\documentclass{aa}  

\usepackage{epsfig,natbib,url,twoopt}
\usepackage{graphicx}
\usepackage{txfonts}
\usepackage{longtable}
\usepackage{supertabular}
\usepackage{lscape}
\usepackage{rotating}
\bibpunct{(}{)}{;}{a}{}{,}

\begin{document} 


\title{The VIMOS Ultra Deep Survey}
\subtitle{The role of HI kinematics and HI column density on the escape of Ly$\alpha$ photons in star-forming galaxies at $2<z<4$}

\author{L. ~Guaita \inst{1}
\and M. ~Talia\inst{2,13}
\and L. ~Pentericci\inst{1}
\and A. ~Verhamme\inst{6}
\and P. ~Cassata\inst{3}
\and B. C. ~Lemaux\inst{5}
\and I.  ~Orlitova\inst{9}
\and B. ~Ribeiro\inst{4}
\and D. ~Schaerer\inst{6}
\and G. ~Zamorani\inst{2} 
\and B.~Garilli\inst{7}
\and V.~Le Brun\inst{4}
\and O.~Le F\`evre\inst{4}
\and D.~Maccagni\inst{7}
\and L. A. M.~Tasca\inst{4}
\and R.~Thomas\inst{3}
\and E.~Vanzella\inst{2}
\and E.~Zucca\inst{2}
\and R. ~Amorin\inst{10,11}
\and S.~Bardelli\inst{2}
\and M. ~Castellano\inst{1} 
\and A.~Grazian\inst{1}
\and N.P.~Hathi\inst{4,8}
\and A.~Koekemoer\inst{8}
\and F. ~Marchi\inst{1}
%
%
%
 \fnmsep\thanks{Based on data obtained with the European Southern Observatory Very Large Telescope, Paranal, Chile, under Large Program 185.A--0791.}
}

\offprints{Lucia Guaita, \email{lucia.guaita@oa-roma.inaf.it}}

\institute{INAF - Osservatorio Astronomico di Roma, Via Frascati 33, 00040 Monteporzio (RM), Italy
\and INAF - Osservatorio Astronomico di Bologna, Via Gobetti 93/3, 40129 Bologna, Italy
\and Instituto de Fisica y Astronom\'ia, Facultad de Ciencias, Universidad de Valpara\'iso, Gran Breta$\rm{\tilde{n}}$a 1111, Playa Ancha, Valpara\'iso Chile
\and Aix Marseille Universit\'e, CNRS, LAM (Laboratoire d'Astrophysique de Marseille) UMR 7326, 13388, Marseille, France
\and Department of Physics, UC Davis, One Shields Avenue, Davis, CA 95616, USA 
\and Geneva Observatory, University of Geneva, 51 Ch. des Maillettes, CH-1290 Versoix, Switzerland
\and INAF--IASF, via Bassini 15, I-20133, Milano, Italy
\and Space Telescope Science Institute, 3700 San Martin Drive, Baltimore, MD 21218, USA
\and Astronomical Institute of the Czech Academy of Sciences, Bo\v cn{\'\i} II/1401, 140 00 Praha 4, Czech Republic
\and Cavendish Laboratory, University of Cambridge, 19 J.J. Thomson Avenue, Cambridge CB3 0HE, UK
\and Kavli Institute for Cosmology, University of Cambridge, Madingley Road, Cambridge, CB30HA, UK
\and Dipartimento di Fisica e Astronomia, Universit\'a di Bologna, Via Gobetti 93/2, 40129 Bologna, Italy
%
}

   \date{Accepted on May 2017}

 
  \abstract
   {}
   {We wish to assess the role of kinematics and neutral hydrogen column density in the escape and distribution of Ly$\alpha$ photons.}
   {We selected a sample of 76 Ly$\alpha$ emitting galaxies from the VIMOS Ultra Deep Survey (VUDS) at $2\leq z\leq4$. 
%
We estimated the velocity of the neutral gas flowing out of the interstellar medium as the velocity offset, $\Delta$v, between the systemic redshift ($z_{\rm sys}$) and the center of low-ionization absorption line systems (LIS). To increase the signal to noise of VUDS spectra, we stacked subsamples defined based on median values of their photometric and spectroscopic properties. We measured the systemic redshift from the rest-frame UV spectroscopic data using the CIII]1908 nebular emission line, and we considered SiII1526 as the highest signal-to-noise LIS line.
We calculated the Ly$\alpha$ peak shift with respect to the $z_{\rm sys}$, the EW(Ly$\alpha$), 
and the Ly$\alpha$ spatial extension, Ext(Ly$\alpha$-C), from the 
profiles 
in the 2D stacked spectra.}
   {The galaxies that are faint in the rest-frame UV continuum, strong in Ly$\alpha$ and CIII], with compact UV morphology, and localized in an underdense environment are characterized by outflow velocities of the order of a few hundreds of km sec$^{-1}$. 
The subsamples with smaller $\Delta$v are characterized by larger Ly$\alpha$ peak shifts, larger Ext(Ly$\alpha$-C), and smaller EW(Ly$\alpha$). In general we find that EW(Ly$\alpha$) anti-correlates with Ext(LyLy$\alpha$-C) and Ly$\alpha$ peak shift.
}
   {We interpret these trends using a radiative-transfer shell model. 
The model predicts that an HI gas with a column density larger than $10^{20}$ cm$^{-2}$ is able to produce Ly$\alpha$ peak shifts larger than $>300$ km sec$^{-1}$. An ISM with this value of N$_{\mathrm {HI}}$ would favour a large amount of scattering events, especially when the medium is static, so it can explain large values of Ext(Ly$\alpha$-C) and small EW(Ly$\alpha$). On the contrary, an ISM with a lower N$_{\mathrm {HI}}$, but large velocity outflows would lead to a Ly$\alpha$ spatial profile peaked at the galaxy center (i.e. low values of Ext(Ly$\alpha$-C)) and to a large EW(Ly$\alpha$), as we see in our data.
Our results and their interpretation via radiative-transfer models tell us that it is possible to use Ly$\alpha$ to study the properties of the HI gas. 
Also, the fact that Ly$\alpha$ emitters are characterized by large $\Delta$v could give hints about their stage of evolution in the sense that they could be experiencing short bursts of star formation that push strong outflows. 
}


\titlerunning{Escape of Ly$\alpha$ photons}
\authorrunning{Guaita L.}

   \maketitle
%

\section{Introduction}



The production of Ly$\alpha$ photons depends on the on-going star formation. However, 
the escape 
depends on a large variety of galaxy properties. It was shown by radiative transfer models, hydrodynamical simulations \citep{V2006,Laursen2011,Orsi2012,Duval2014,Gronke2016}, and observed in high- and low-redshift galaxies \citep{Erb2014,Rivera-Thorsen2015,Henry2015,Trainor2015} that interstellar medium (ISM) dust is able to absorb Ly$\alpha$ photons and neutral hydrogen (HI) is able to scatter them. The emerging Ly$\alpha$ line profile can present two peaks \citep{Neufeld1990,Dijkstra2006,V2006}. The dominant red peak can be detected even in low-resolution spectra and its shift with respect to the systemic redshift is related to the neutral hydrogen column density (N$_{\mathrm {HI}}$) and gas kinematics such as stellar outflows. 

In medium- and high-resolution spectra, it is also possible to detect a blue peak. 
Radiative transfer models predict that, for simple geometries, the intensity of the blue peak and the separation between red and blue peaks are also connected with the HI column density \citep{Verhamme2015} and  dust content. 
\citet{Hashimoto2015} estimated the N$_{\mathrm {HI}}$ in a sample of $z\sim2.2$ galaxies by comparing the Ly$\alpha$ profiles derived from the 1D spectra with the models described in \citet{Schaerer2011}. For the Ly$\alpha$ emitting galaxies that showed a significant blue peak, these authors estimated N$_{\mathrm {HI}}$ lower than $10^{19}$ cm$^{-2}$. 
Also, they measured an average velocity offset of the red peak with respect to the systemic redshift of the order of 170 km sec$^{-1}$ and explained this value by  N$_{\mathrm {HI}}$ as low as $10^{18.9}$ cm$^{-2}$ \citep[see also][]{Verhamme2017}. 
\citet{Shibuya2014} investigated the role of the interstellar medium HI kinematics on the escape of Ly$\alpha$ photons in a similar sample. They inferred the velocity of stellar outflows from the shift of the low-ionization absorption lines with respect to the systemic redshift (redshift of optical emission lines). They did not find any clear correlation between outflow velocity and Ly$\alpha$ equivalent width and proposed that low N$_{\mathrm {HI}}$ can make a galaxy appear as a strong Ly$\alpha$ emitter. However, it is difficult to disentangle the effect of kinematics and N$_{\mathrm {HI}}$ because a proper kinematics measurement implies a very good knowledge of the galaxy systemic redshift. In addition to this, complicated kinematics in the interstellar and also in the intergalactic medium can influence the Ly$\alpha$ photon emission. For instance, the data studied in \citet{Smit2017} are consistent with a picture in which outflows of different velocities can shape the Ly$\alpha$ emission line at different wavelengths starting from the main peak to the wings.

The propagation of Ly$\alpha$ photons in the circum-galactic medium is also related to the HI properties and its scattering capability. The total size of the Ly$\alpha$ emission was observed to be generally more extended than the rest-frame UV continuum \citep{Steidel2011,Hayes2013,Momose2014,Guaita2015,Wisotzki2016}. In a sample of local star-forming galaxies, \citet{Pardy2014} and \citet{Hayes2014} found that an HI gas characterized by N$_{\mathrm {HI}}$ $>10^{20}$ cm$^{-2}$ can efficiently scatter Ly$\alpha$ photons and produce the most extended Ly$\alpha$ emissions. They also showed that the Ly$\alpha$ equivalent width anti-correlates with the HI total mass. 

\citet{Matsuda2012} proposed that the Ly$\alpha$ spatial extension strongly depends on environment. In fact, 
they found that 
the spatial extension of Ly$\alpha$ increases as their surface density increases and that the highest values are observed in proto-clusters \citep[see also][]{Steidel2011}. 
Furthermore, \citet{Momose2016} studied the Ly$\alpha$ spatial extension as a function of physical properties. They found that the largest Ly$\alpha$ emission sizes are observed for the galaxies with the smallest Ly$\alpha$ equivalent width and brightest UV luminosity, which are possibly the most massive 
\citep[see also][]{Zheng2016}.

%
In the past 10 years, the physical properties of the Ly$\alpha$ emitting galaxies and in particular of the LAEs 
\citep[Ly$\alpha$ emitting galaxies with EW(Ly$\alpha)>20$ {\AA} as defined for example in][]{Gronwall:2007} have been inferred from the spectral energy distribution analysis and from the multi-wavelength images and spectra.
The LAEs tend to occupy the low-mass end of the star-forming galaxy mass distribution; they are compact in the rest-frame UV, characterized by low dust content, low metallicity, high specific star formation rates, and tend to be experiencing bursts of star formation \citep[e.g.][]{Fin2011,Nakajima2012a,Hashimoto2013,Shibuya2013, Vargas2014,Hagen2014}. However, Ly$\alpha$ in emission is also observed for galaxies with a moderate amount of dust \citep[e.g.][]{kornei2010,Guaita2011,Hathi2016}. Also, there is evidence that the largest equivalent width LAEs present high-ionization state and low oxygen abundances \citep{Erb2010,Nakajima2012b,Trainor2016,Nakajima2016,Amorin2017}. In fact, high-ionization absorption, such as the CIV$\lambda \lambda$1548,1550 doublet, and strong nebular emission lines, such as the collisionally excited CIII$\lambda \lambda$1907,1909 doublet, are detected in their spectra. 

%
\citet{Stark2014} suggested a correlation between the equivalent width of CIII$\lambda \lambda$1907,1909 and Ly$\alpha$. They studied a sample of 17 strongly lensed galaxies at $z\sim 2$ characterized by EW(CIII]1908) up to 15 {\AA} and EW(Ly$\alpha$) up to 150 {\AA} and found that they are characterized by low stellar masses, 
 low metallicity, 
 and specific star formation rates more than 10 times those of typical $z\sim2-3$ UV-selected star-forming galaxies.
More recently, similar results were found by \citet{Amorin2017} (hereinafter A17) for a sample of 10 low-mass 
 galaxies at $z\sim3$, selected by their strong Ly$\alpha$ (EW(Ly$\alpha) > 45$ {\AA}) in the VIMOS Ultra-Deep Survey \citep{LeFevre2015}. 
 While their properties closely resemble those typically found in galaxies at higher redshift \citep[$z>6$; e.g.][]{Smit2014,Stark2017}, the above studies suggest that low-mass galaxies showing strong CIII] and Ly$\alpha$ emission might be experiencing a vigorous phase of rapid galaxy growth \citep[e.g.][]{Gawiser:2007}. 
%
%
However, other studies just show correlations with some scatter  \citep[e.g.][]{Rigby2015}.
The model described in \citet{Jaskot2016}, for example, shows a weak trend between the equivalent widths of CIII] and Ly$\alpha$. The scatter between these two quantities depends on the ionization parameter and metallicity of the starburst regions.

Using rest-frame UV spectroscopy of galaxies with both Ly$\alpha$ and CIII] in emission, we may focus on extreme cases of Ly$\alpha$ emission, but we can set the systemic redshift just looking at the same spectrum. Knowing the systemic redshift allows us to derive information on the kinematics of the ISM, obtained from low-ionization absorption lines, which trace the HI gas. 
Therefore, we can 
disentangle the effect of kinematics and HI column density on the shape of the Ly$\alpha$ emission line and its spatial extension. 
With a large sample of Ly$\alpha$+CIII] emitters from the VIMOS Ultra-Deep Survey \citep[VUDS\footnote{http://cesam.lam.fr/vuds/};][]{LeFevre2015, Tasca2016},
we aim to study and characterize the spatial versus spectral escape of the Ly$\alpha$ emission from galaxies to investigate further if scattering is the main powering mechanism of the extended Ly$\alpha$ emission around galaxies.
To do this, we present the Ly$\alpha$ equivalent width, the shape of the Ly$\alpha$ emission line, and the Ly$\alpha$ spatial extension in stacked spectra. 

The paper is organized as follows. In Sect. 2 we present the spectroscopic data used in this analysis; in Sect. 3 we describe the method used to estimate the systemic redshift, 
generate stacked spectra, 
and measure kinematics features; in Sect. 4 and 5 we show and discuss our results. We summarize our work in Sect. 7. 
Throughout the paper, we use AB magnitudes and air wavelengths, the equivalent widths are expressed in the rest-frame system, and we adopt a standard cosmology. 


\section{Spectroscopic data}
\label{sec:Data}

%

The spectra of the galaxies we analyse in this work are part of the VUDS, which is a
deep spectroscopic survey of 10000 star-forming galaxies (SFGs) performed with the
VIsible MultiObject Spectrograph (VIMOS) instrument at the Very Large Telescope. 
The survey was designed to provide a complete census of high-redshift galaxies with the scope of studying the history of the global star formation, the build up of the mass function, and very young objects among many others. The VIMOS spectra are obtained with the LRBLUE and LRRED-grism settings in low-resolution mode (R $\sim300$).  
The survey covers three extragalactic fields: the COSMOS field, the extended Chandra Deep Field South (ECDF-S), and the VVDS-02h field. Each spectrum is the result of 14 hours of integration in a wavelength range between $\sim$4000 and $\sim$9500 {\AA}, which allows the study of SFGs at $2\leq z\leq7$, and is calibrated to air wavelength (for reference $\lambda_{\rm vacuum}$ ([CIII]) $=1907.71$ {\AA} and $\lambda_{\rm air}$ ([CIII]) $=1907.07$ {\AA}). Among the SFGs, we focused on the Ly$\alpha$ emitting galaxies \citep[EW(Ly$\alpha)>0$ {\AA}, based on the EW measurements from][]{Cassata2015} with reliable redshift (flags 3, 4, corresponding to a probability greater than 95\% for the redshift to be correct) 
at $2\leq z\leq4$. 

We excluded the sources detected in the most recent Xray surveys, which are thought to be active galactic nuclei (AGNs). For the COSMOS field, we cross-matched the VUDS Ly$\alpha$ emitting galaxies with the catalogue presented in \citet{Civano2016}. We also double checked any possible Xray emission considering the catalogues by \citet{Civano2012}, \citet{Kim2007Champ}, and \citet{K2014Civano}. For ECDF-S, we considered  the X-ray catalogues from \citet{Xue2011} and from the 3XMM\_DR4 survey\footnote{http://xmmssc-www.star.le.ac.uk/Catalogue/3XMM-DR4/}. We also removed from our sample the sources from the AGN lists by \citet{Fiore2012} and by \citet{Xue2011}. For the VVDS-02h field, we excluded any source found in the XMDS/VVDS 4$\sigma$ \citep{Chiappetti2005} and in the 3XMM\_DR4 survey catalogues.

 The wavelength range of the spectra at $2\leq z\leq4$ includes Ly$\alpha$, low-ionization interstellar absorption line systems (LIS), such as SiII1260, CII1334, SiII1526, 
and spectral features that trace the systemic redshift ($z_{\rm sys}$). The LIS lines are produced in the ISM by the absorption of the stellar UV radiation and trace the kinematics of the HI gas \citep[e.g.][]{Talia2012}. 
The photospheric stellar absorption lines, such as OIV1343, SiIII1417, and SV1500 \citep[e.g.][]{Talia2012}, are the most commonly used systemic-redshift sensitive features.
However, these lines tend to be very weak and do not reach enough signal to noise to provide the $z_{\rm sys}$ of our individual spectra. Nebular emission lines, such as the CIII$\lambda \lambda$1907,1909 doublet, are produced around massive stars in HII regions and can also trace the $z_{\rm sys}$. Talia et al. (in prep) showed concordance between $z_{\rm sys}$ obtained using stellar photospheric lines and that obtained from CIII] for the star-forming galaxies in VUDS.

In the individual spectra of the galaxies in our sample, 
the CIII doublet is blended and we refer to it as CIII]1908 \citep{Stark2014}. Even if we are limited by the low resolution, the CIII]1908 emission line provides the most reliable systemic redshift when its (integrated flux) signal to noise (SN) is larger than 3. 
We excluded the cases in which CIII] is blended to any OH line or is close to any artificial defect, which are situations that increase the noise locally (see Sec. \ref{sec:zsys}). 

 Starting from 1070 Ly$\alpha$ emitting galaxies at $2<z<4$, we kept 79 showing SN(CIII]1908) $>3$. We then excluded three sources with UV-line ratios consistent with an AGN ionization. We measured CIV1550/CIII]1908, CIV1550/HeII1640 and we referred to the figure A.2 in \citet{Feltre2016} for the exclusion. This figure shows CIV1550/HeII1640 versus CIV1550/CIII]1908 and highlights the values of those ratios typical of star-forming galaxies and AGN.
Finally, we compiled a sample of 76 galaxies. Among these galaxies,  about two-thirds are characterized by Ly$\alpha$ equivalent width (EW) larger than 20 {\AA}, the value commonly used to define Ly$\alpha$ emitters (LAEs). 

\begin{figure*}
\centering
\includegraphics[width=70mm]{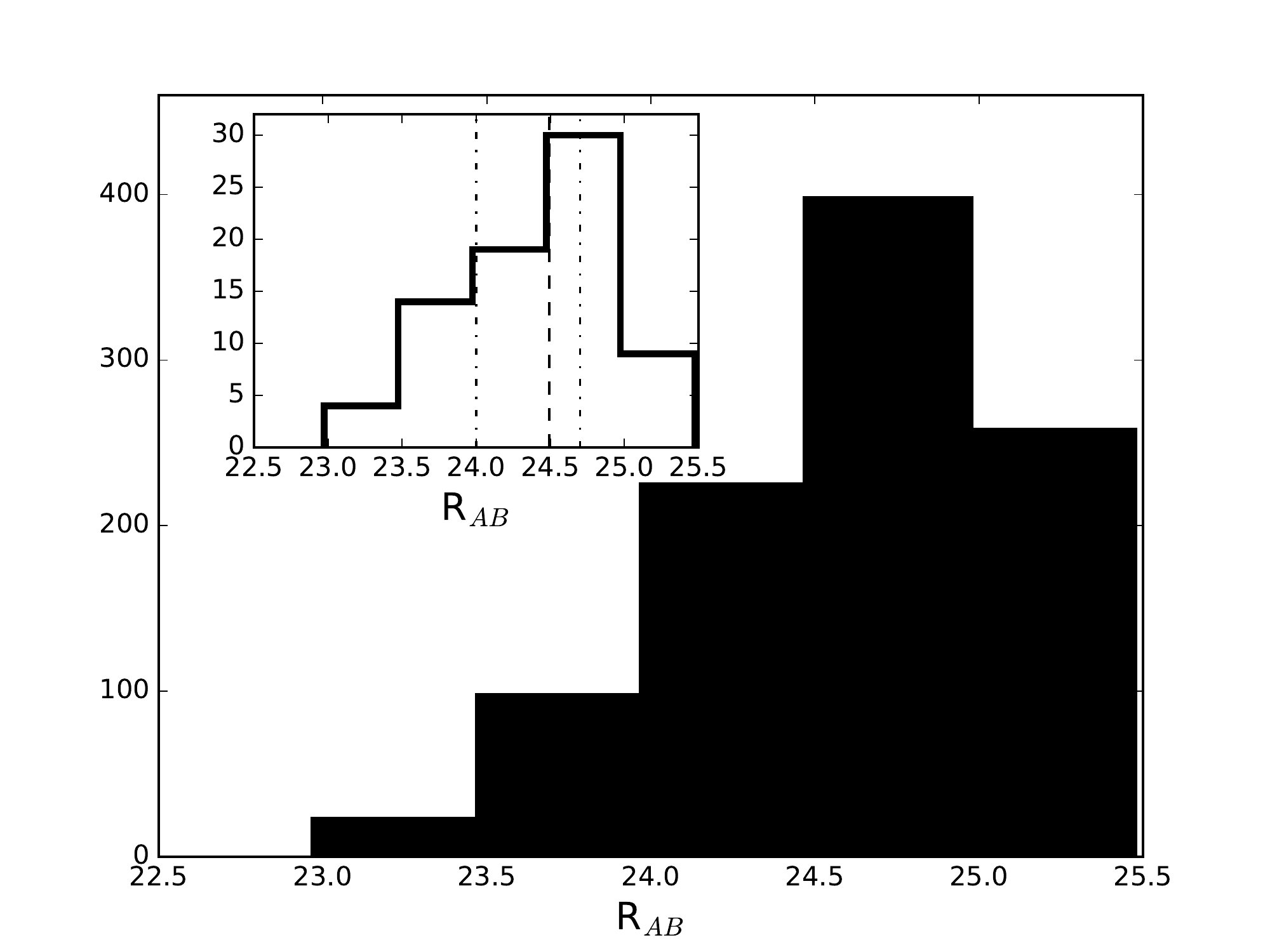}
\includegraphics[width=70mm]{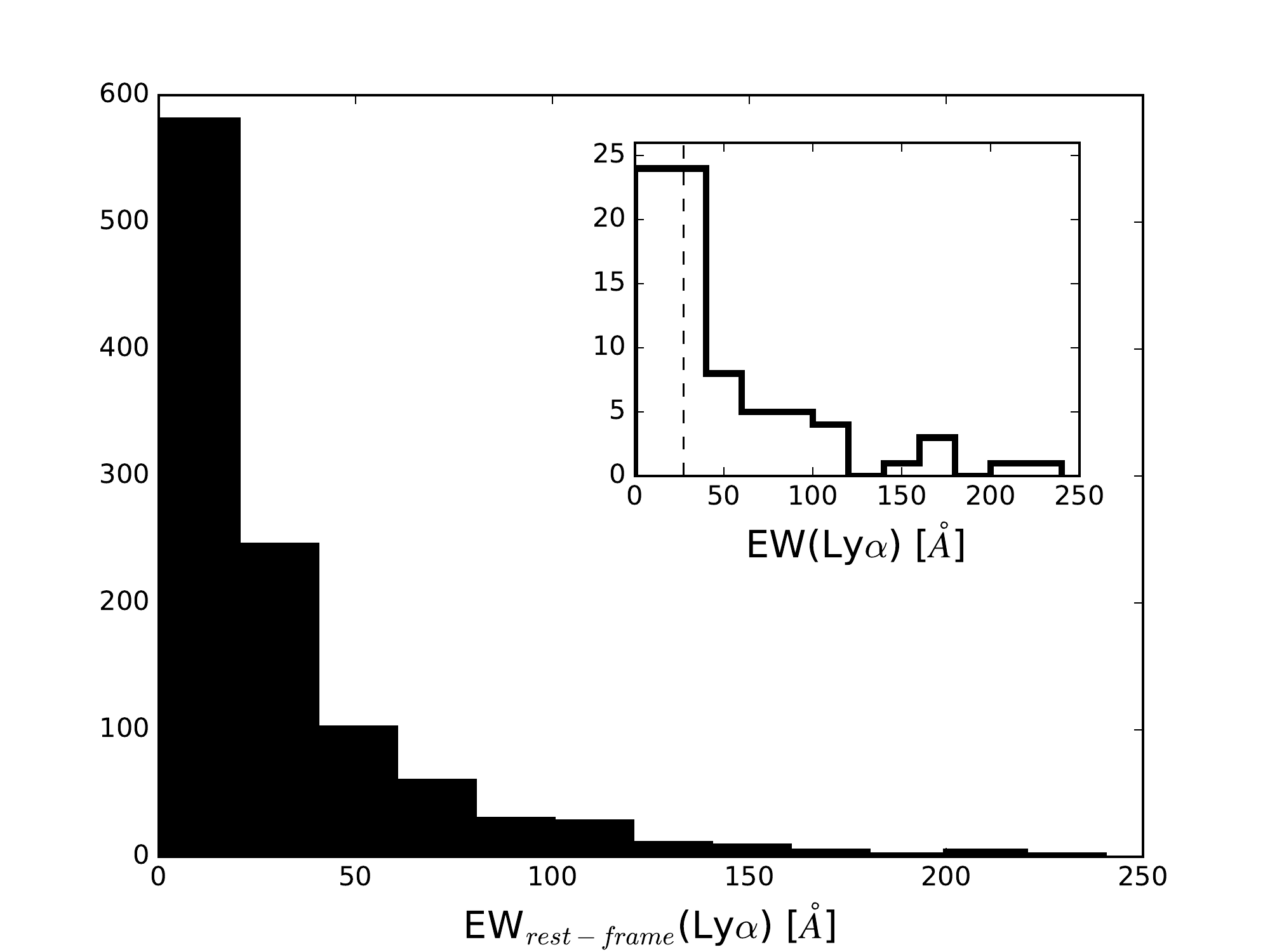}
\includegraphics[width=70mm]{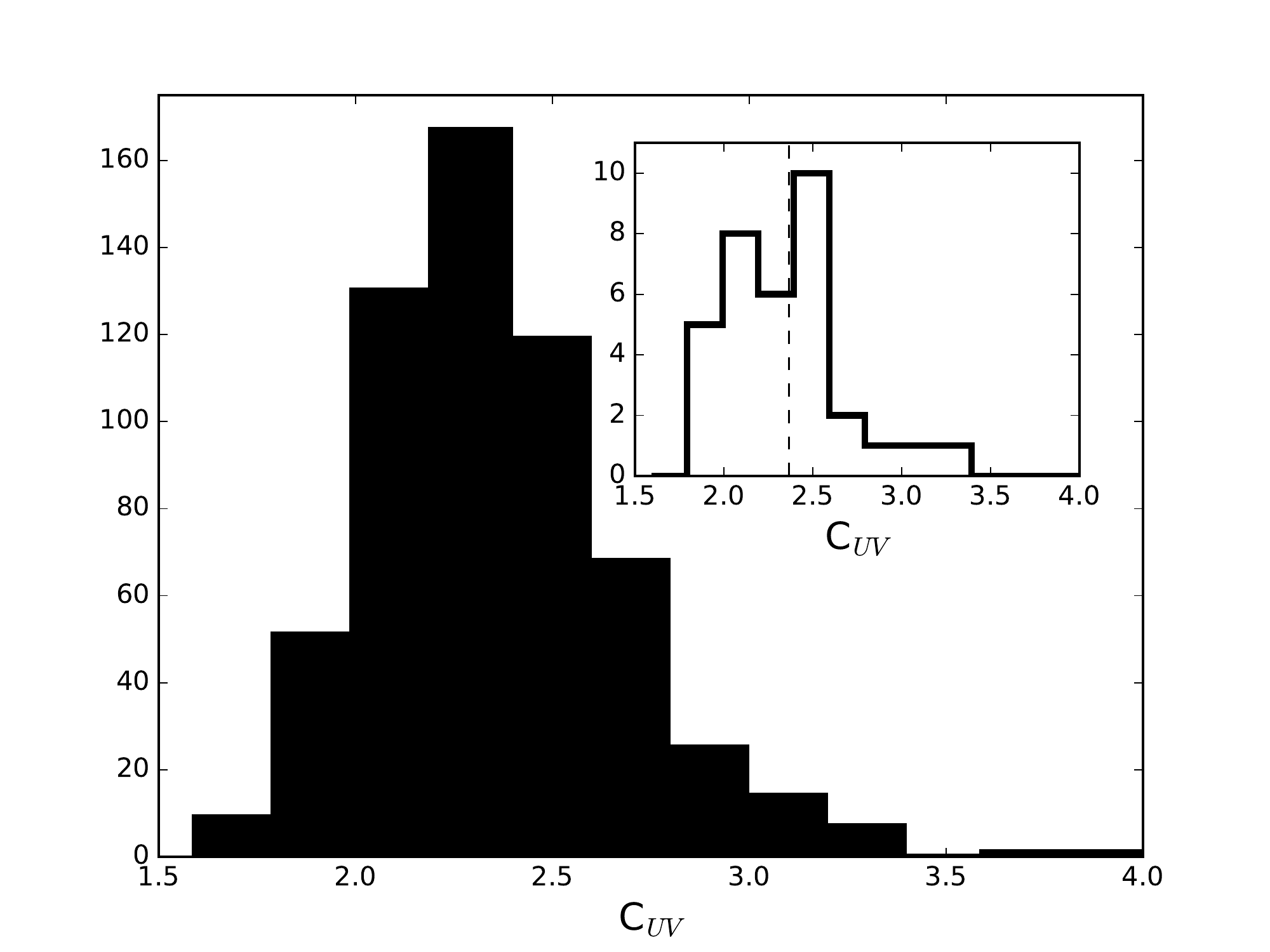}
\includegraphics[width=70mm]{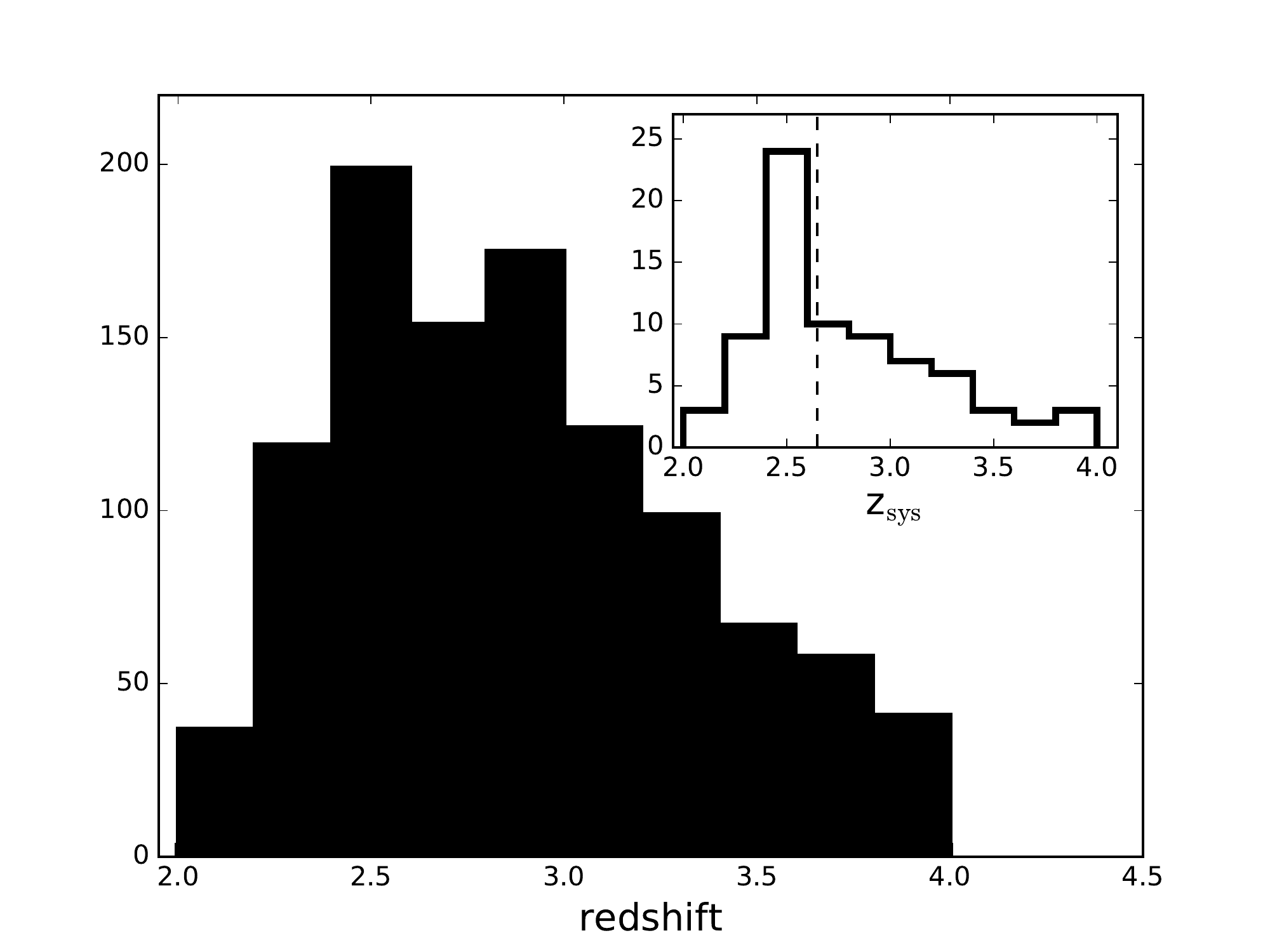}
\caption{Distribution of $R$ magnitude ($top ~left$), EW(Ly$\alpha$) ($top ~right$), 
rest-frame UV concentration ($bottom ~left$), 
redshift ($bottom ~right$) of our sample of 76 sources (empty histograms in the inserts). Filled histograms correspond to the distributions of the 1070 Ly$\alpha$ emitting galaxies at $2<z<4$. 
The vertical dashed lines indicate the median value of each histogram. In the $top ~left$ panel, the dot-dashed vertical lines indicate the median $R$ values of the two bins obtained, separating the sample with respect to the median of the 76-galaxy distribution.}
\label{fig:prop}
\end{figure*}

In Fig. \ref{fig:prop}, we show the distributions of the properties of the 76 galaxies and of the initial 1070 sources. 
Our sample is not biased with respect to the initial sample in terms of physical properties, such as rest-frame UV magnitude, concentration, and redshift, except for EW(Ly$\alpha$) and stellar mass.  
The median EW(Ly$\alpha$) value of our sample is significantly larger than that of the 1070 sources. 
This is not surprising since we are selecting Ly$\alpha$ emitting galaxies that are also CIII] emitters with SN(CIII]1908) $>3$ and since earlier studies found, in general, a correlation between the CIII] and Ly$\alpha$ equivalent widths \citep{Stark2014}. 

We considered the physical parameters obtained with the Le Phare SED fitting code \citep{Arnouts1999}, which was applied to the VUDS data as outlined in \citet{Tasca2016}. 
By starting from the best available multiwavelength photometry and fixing VUDS spectroscopic redshift, they applied seven different \citet{Bruzual:2003} models to obtain the best-fit stellar mass, star formation rate (SFR), age, and dust reddening. The models encompass star formation histories characterized by an exponentially decaying mode and delayed exponentially decaying modes. 
 A \citet{chabrier2003} initial mass function is used and nebular emission lines are added to the stellar templates \citep{Ilbert2009}.
The Ly$\alpha$+CIII] emitters of our sample are characterized by stellar masses of log(M$_{*}$/M$_{\odot}$) = [8.4-10.9] with a median value of log(M$_{*}$/M$_{\odot}$) = 9.5, dust reddening E(B-V) = [0.0-0.3] with a median value of 0.1, and (uncorrected) SFR = 4-200 M$_{\odot}$ yr$^{-1}$ , with a median value of 20 M$_{\odot}$ yr$^{-1}$. 

\begin{figure*}
\centering
\includegraphics[width=200mm]{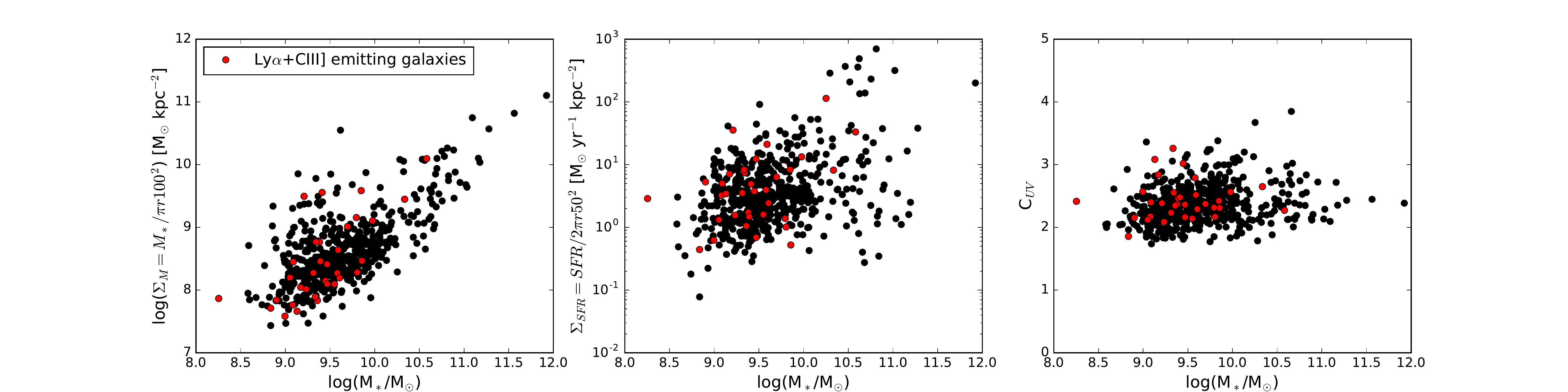} 
\caption{Stellar mass surface density ($left$), star formation rate surface density ($middle$), and rest-frame UV concentration ($right$) as a function of stellar mass. To calculate the stellar mass surface density, we used the total-light radius, r100, while to calculate the star formation rate surface density, we used the half-light radius, r50, according to the definitions in \citet{Ribeiro2016}. The size measurements can be performed for a limited number of sources for which the HST coverage is available. Black dots represent the Ly$\alpha$ emitting galaxies at $2<z<4$, while red dots correspond to those with SN(CIII]1908) $>3$. 
}
\label{fig:comparative}
\end{figure*}

In Fig. \ref{fig:comparative} we show stellar mass surface density, star formation rate surface density, and concentration versus stellar mass for the 1070 sources and for our 76 Ly$\alpha$+CIII] emitters. 
The median stellar mass of our sample 
is 30\% smaller than that of the initial sample 
and it is not characterized by extreme values of SFR and star formation rate surface density.
Compared with the stellar mass surface density versus stellar mass of all the galaxies in VUDS \citep[][Fig. 2 and 19]{Ribeiro2016}, our sample occupies the low-mass end. 

To summarize, our sample is composed of galaxies that tend to be faint and moderately concentrated in the rest-frame UV continuum. 
These galaxies are characterized by median rest-frame EW(Ly$\alpha)=27$ {\AA}, EW(CIII]1908) $=7.9$ {\AA}, and a median stellar mass of $2.9 \times10^{9}$ M$_{\odot}$ 
(Fig. \ref{fig:prop} and Table \ref{tab:sub}). 

%


\section{Method}
\label{sec:Method}

Since the escape and the large-scale propagation of Ly$\alpha$ photons depend on the ISM kinematics, 
we wish to estimate kinematics features and its dependency with Ly$\alpha$ equivalent width, Ly$\alpha$ line shape, and Ly$\alpha$ spatial extension.

We first measured the systemic redshift of each Ly$\alpha$+CIII] emitting galaxy (Sec. \ref{sec:zsys}) and we defined six pairs of subsamples (listed in Table \ref{tab:sub} and described in Sec. \ref{sec:subsample}), based on physical and spectral properties, i.e. rest-frame UV magnitude, Ly$\alpha$ and CIII]1908 equivalent width, rest-frame UV concentration, environment density, and stellar mass; we stacked these pairs to increase the signal to noise of the individual spectra (Sec. \ref{sec:stack}). Then, we considered the LIS lines 
that trace the neutral gas to infer the HI kinematics (such as the velocity of stellar inflows/outflows). We quantified the kinematics by evaluating the offset between the systemic redshift and the central wavelength of the highest signal-to-noise LIS (Sec. \ref{sec:kin}).
%

\subsection{Systemic redshift}
\label{sec:zsys}

\begin{figure}
\centering
\includegraphics[width=80mm]{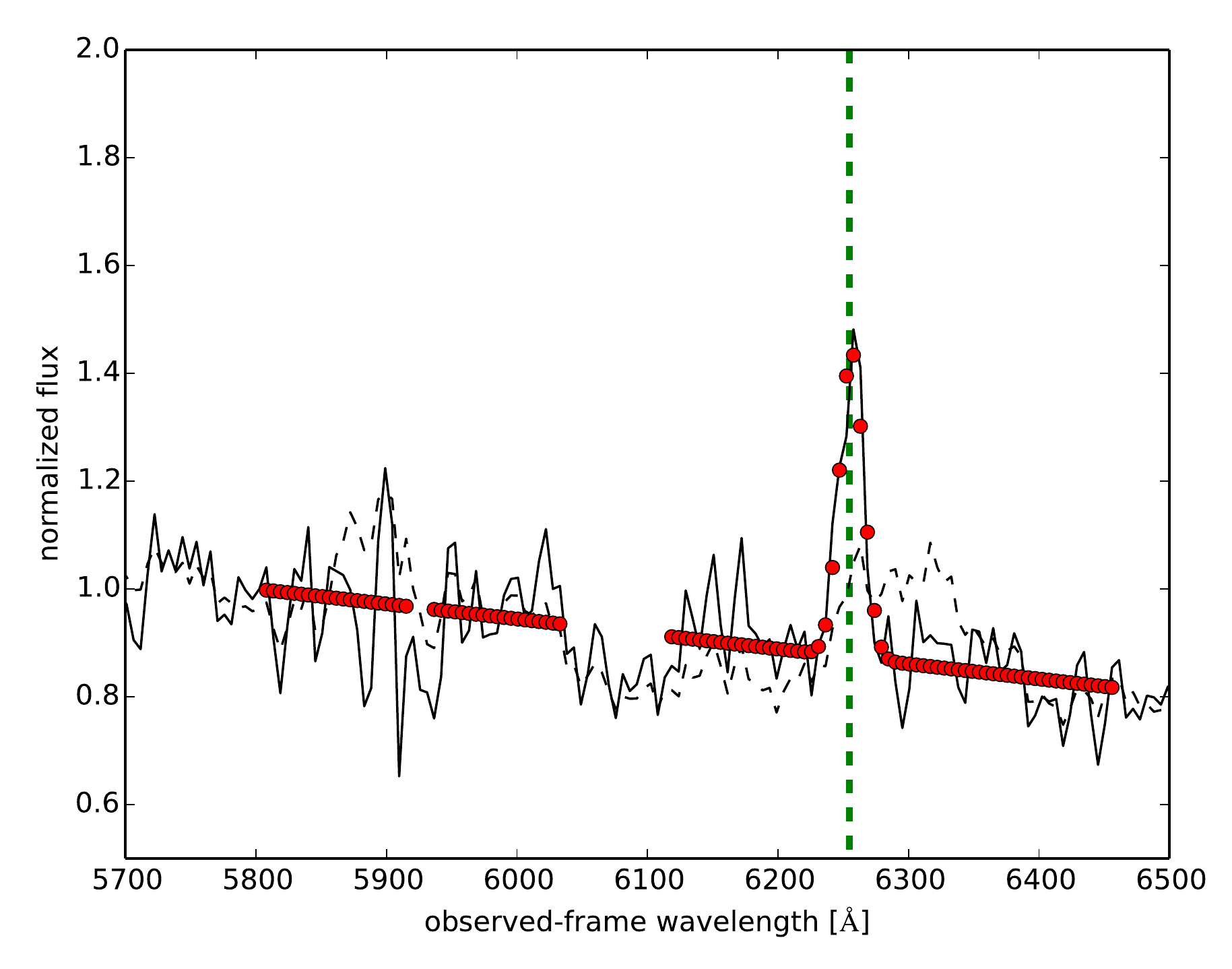} 
\caption{Region of the spectrum covering the CIII]1908 emission line and the continuum around it for the source \#510838687. The air observed-frame (noise) spectrum is presented as a solid (dashed) line. It is normalized to the continuum value. 
The vertical green line indicates the (air) wavelength of CIII] at VUDS redshift. }
\label{fig:zsys}
\end{figure}

%
The rest-frame UV continuum of the Ly$\alpha$ emitting galaxies in our sample is faint with a median $R$ magnitude of 24.5. Stellar photospheric absorption lines, which are intrinsically weak features, have a signal to noise that is too low and cannot be recognized in individual spectra.
Therefore, we estimated the systemic redshift ($z_{\rm sys}$) from the CIII]1908 (Fig. \ref{fig:zsys}). 
To estimate $z_{\rm sys}$, we started with the VUDS redshift as a guess, and we fit together the CIII]1908 emission line and the continuum around it. We used the $optimize.leastsq$ python function \citep[as applied also in][]{Guaita2013} by assuming a Gaussian profile, that the continuum has a well-defined slope (linear fit),  
 and including the VUDS noise spectrum in the minimization procedure. This noise contains the information of the observation and the instrumental uncertainty. In fitting the continuum, we excluded the wavelength regions occupied by absorption lines that could alter the slope. 
We derived the galaxy systemic redshift from the best-fit mean of the Gaussian profile. 

\subsection{Definition of the subsamples}
\label{sec:subsample}

From the VUDS spectra, we can compute three complementary Ly$\alpha$ quantities: the equivalent width EW(Ly$\alpha$), the Ly$\alpha$ peak shift compared to systemic redshift, and the spatial extension of the Ly$\alpha$ line in the direction perpendicular to the slit (see the definitions below). The last two quantities are not measurable with enough accuracy on individual spectra, but on stacks only, as described later. In order to investigate which physical parameter, if any, drives the escape of Ly$\alpha$ radiation from galaxies, we computed several physical quantities for our galaxies. Then, we separated the sample of 76 galaxies in halves, according to the median values of $R$ magnitude (representing the rest-frame UV at $2<z<4$), EW(Ly$\alpha$), 
EW(CIII]1908), 
 and stellar mass 
as shown in Table \ref{tab:sub}. 
%
%

The equivalent widths of CIII]1908 are consistent with the values typical of metal-poor galaxies (A17). 
\begin{table*} 
\centering
\caption{Characteristics of the sub-sample stacks}
\label{tab:sub}
\scalebox{0.78}{
\begin{tabular}{|c|c|c|c|c|c|c|c|c|c|c|c|c|} 
\hline
%
%
%
sub-sample & N & $z_{\rm sys ~median}$ & R$_{\rm median}$ & SFR$_{\rm median}$ & $\Sigma_{\rm SFR ~median}$ & M$_{\rm * ~ median}$ &  $\Delta v$(SiII) & EW(SiII) & EW(CIII]) & EW(Ly$\alpha$) & Ext(Ly$\alpha$-C) & Ly$\alpha$ peak shift \\
property& & & AB & M$_{\odot}$yr$^{-1}$ & M$_{\odot}$yr$^{-1}$ kpc$^{-2}$ &  1E+9 M$_{\odot}$    &   km sec$^{-1}$ & {\AA}& {\AA}&{\AA} & kpc &km sec$^{-1}$ \\
(1) & (2) &(3)  & (4)  & (5)  & (6)  &(7) &(8) & (9) &(10)  &(11) & (12) & (13) \\ 
\hline
\hline
R$\geq24.5$ &  38 &   2.9  & 24.7$\pm$0.1 & 11.0$\pm$3.3 & 4.4$\pm$1.1  & 2.8$\pm$2.4 & -280$\pm$100 & -1.6$\pm$0.5 & 13.3$\pm$1.8  & 58.4$\pm$9.0  &    5.3$\pm$0.6  & 173$\pm$62 \\
R$<24.5$ &  38 &   2.5  & 24.0$\pm$0.1 & 28.8$\pm$8.6 & 2.8$\pm$1.7   & 3.1$\pm$2.5 &  -20$\pm$80  & -0.9$\pm$0.3  & 6.2$\pm$0.9  & 22.7$\pm$3.8  &  6.0$\pm$0.6  & 428$\pm$110 \\
\hline
EW(Ly$\alpha)\geq27${\AA} &  38 &   2.7  & 24.6$\pm$0.1 & 14.0$\pm$4.6 & 4.5$\pm$1.5  & 3.1$\pm$3.1 & -360$\pm$130  & -1.2$\pm$0.4  &11.8$\pm$2.2  & 71.3$\pm$10.0  &   5.3$\pm$0.4  & 196$\pm$60 \\
EW(Ly$\alpha)<27${\AA} &  38 &   2.6  &  24.3$\pm$0.1 & 19.5$\pm$8.7 & 2.9$\pm$0.5  & 2.8$\pm$1.6 & -30$\pm$60  & -1.6$\pm$0.6  & 7.6$\pm$1.0  & 9.7$\pm$1.3  &  6.4$\pm$0.6  & 510$\pm$98 \\
\hline
EW(CIII]) $\geq7.9${\AA} &  38 &   2.9  & 24.7$\pm$0.1 & 12.7$\pm$8.3 & 4.1$\pm$1.1  & 3.0$\pm$3.3 & -430$\pm$180  & -1.2$\pm$0.7  & 15.2$\pm$2.0  & 58.2$\pm$11.6  &  4.8$\pm$0.5  & 207$\pm$80 \\
EW(CIII]) $<7.9${\AA} &  38 &   2.5  & 24.2$\pm$0.1 & 20.1$\pm$5.6 & 3.3$\pm$1.6 & 2.7$\pm$0.8  & -30$\pm$50  & -1.8$\pm$0.3  & 4.6$\pm$0.2  & 22.0$\pm$4.3  &   6.3$\pm$0.5  & 307$\pm$58 \\
\hline
C$_{UV}\geq2.4$ &  17 &   2.7  & 24.7$\pm$0.1 & 11.7$\pm$3.4 & 4.1$\pm$1.2   & 2.9$\pm$2.1  &  -470$\pm$200  & -2.1$\pm$1.3   & 13.3$\pm$3.6  & 94.2$\pm$18.3  &   4.7$\pm$0.9  & 93$\pm$72 \\
C$_{UV}<2.4$ &  17 &   2.7  & 24.5$\pm$0.1 & 20.5$\pm$4.0 & 3.0$\pm$1.5 & 2.9$\pm$2.4  & -80$\pm$120  &-2.0$\pm$0.5  & 9.1$\pm$1.8  & 16.0$\pm$4.2  &   4.1$\pm$1.4  & 325$\pm$140 \\
\hline
Over &  50 &   2.7  & 24.5$\pm$0.1 & 17.5$\pm$4.9 & 3.7$\pm$1.4  & 3.2$\pm$1.8  &  -70$\pm$60  & -1.1$\pm$0.3  & 8.8$\pm$1.0  & 32.3$\pm$5.8  &   5.7$\pm$0.5  & 284$\pm$76 \\
Under &  26 &   2.6  & 24.4$\pm$0.1 & 17.6$\pm$11.2 & 3.5$\pm$0.6  & 2.7$\pm$3.7 & -400$\pm$170  & -1.8$\pm$0.8  &11.5$\pm$3.0  & 58.4$\pm$13.8  &   5.0$\pm$0.8  & 160$\pm$85 \\
\hline
M$_{*} \geq$ 2.9E+9 M$_{\odot}$ &  38 &   3.0  & 24.3$\pm$0.1 & 21.9$\pm$9.0 & 4.3$\pm$1.6  & 5.9$\pm$3.1 & -270$\pm$120  & -1.2$\pm$0.3  &  11.0$\pm$2.0  & 38.7$\pm$9.5  &  5.0$\pm$0.6  & 137$\pm$59 \\
M$_{*} <$ 2.9E+9 M$_{\odot}$ &  38 &   2.6  & 24.5$\pm$0.1 & 14.7$\pm$3.3 & 3.1$\pm$0.9  & 1.9$\pm$0.2 & -63$\pm$110  & -1.4$\pm$0.5  &  8.0$\pm$0.8  & 41.0$\pm$7.6  & 5.8$\pm$0.5  & 322$\pm$69 \\
\hline
\hline
\end{tabular}
}
\tablefoot{
The columns correspond to (1) the properties that define the sub-samples: $R$ magnitude, Ly$\alpha$ and CIII]1909 equivalent widths, rest-frame UV concentration, 
field density, and stellar mass of the galaxies in our sample; (2) the number of sources in each sub-sample. Some of the sources in the sub-samples selected based on different properties overlap. The concentration is estimated only when we have HST coverage; 
for the sources in each sub-sample we report (3) the median systemic redshift, (4) the median $R$-band magnitude, (5) the median star-formation rate, 
obtained from the SED fitting, (6) the median star formation rate density calculated from the SFR and from the PSF-corrected half-light-radius measured in the HST $I$ broad-band images ($\Sigma_{SFR}=$SFR/2 $\pi$ r$_{1/2}^2$), and (7) the median stellar mass also obtained from the SED fitting. We report the error bars on the median values. It is up to 0.1 for $R$, of the order of 5 for SFR, 0.5-1.7 for $\Sigma_{SFR}$, 0.2-3.7 for M$_{*}$. 
Then, we show six quantities calculated in the stacks of the sub-samples:  (8) the velocity offset between CIII] and SiII1526, (9) the rest-frame equivalent width of the SiII1526 absorption line, (10) the rest-frame equivalent width of CIII]1908, (11) the rest-frame equivalent width of Ly$\alpha$, (12) the 2D Ly$\alpha$ spatial extension, (13) the Ly$\alpha$ peak shift. 
}
\end{table*}

The equivalent widths of Ly$\alpha$ and CIII]1908 were calculated in the 1D spectra from the flux integrated within the Gaussian fits of the lines and the fitted level of the local continua. 
From our total sample of 76 sources, 34 are located in the fields with HST coverage \citep{Koekemoer2007,Koekemoer2011,Grogin2011}; we consider the rest-frame UV light concentration, C$_{UV}$, measure in \citet{Ribeiro2016} (equations 6, 7, 18, and 21). We generated two subsets, separating the 34 galaxies in halves according to the median C$_{UV}$ value (2.4).
A $Monte ~Carlo$ Voronoi tessellation technique was applied to the VUDS fields to estimate local (number) densities of galaxies \citep[][and Lemaux et al., in prep, where the technique is adapted for VUDS]{Lemaux2016}. We defined the delineation point for overdense and underdense regions as being the median density of a sample of VUDS galaxies in the redshift range $2 < z < 4$ 
matched in stellar mass to the Ly$\alpha$ emitting galaxy sample presented in this study.   
According to this density definition, we found that 26 out of 76 Ly$\alpha$ emitting galaxies are located in underdense and 50 in overdense regions. We generated the stacks of the 26 and 50 sources as shown in Table \ref{tab:sub}.

\subsection{Stacking procedure}
\label{sec:stack}

To generate the stacked 1D spectra, we normalized each observed-frame spectrum to the value of its continuum at 1650-1850 {\AA}, a wavelength region free of strong absorption lines \citep{Talia2012}, and we obtained the 
rest-frame spectra using their $z_{\rm sys}$. We took the median of the systemic redshifts within each subsample as the representative of the subsample stack. The range in $z_{\rm sys}$ within each subsample produces a range in wavelength steps of less than 0.5  {\AA} in the rest frame (Fig. \ref{fig:prop}).  
Therefore, we resample each spectrum to a wavelength step of 
\begin{equation}  
step_{\mathrm \lambda}=\frac{5.355}{1+median(z_{\rm sys})}, 
\end{equation}
where 5.355 is the nominal VIMOS\_LR {\AA}/pixel scale (of the observed-frame spectra) and $median$($z_{\rm sys}$) is the median $z_{\rm sys}$ of a certain subsample. 
Then, we stack the normalized rest-frame spectra by performing an average combination. The whole spectrum of each subset is presented in Appendix A. 
The noise of the stacked spectrum is created from the square root of the sum of the squares of the VUDS noise spectra of the N sources in a subsample, \\
\begin{equation}
noise_{stack}=\frac{\sqrt{(\sum_i^N(noise_i^2)}}{N}. 
\end{equation}
 In Fig. \ref{stackedspectra1} and \ref{stackedspectra2}, we show the stacked 1D spectra of each subsample and we indicate emission and absorption features with vertical lines. The blue dashed, red dashed, and green solid lines indicate Ly$\alpha$, HeII, and CIII], respectively. The black dashed lines indicate LIS and the dashed yellow stellar absorption lines. The stellar features are within the continuum noise. Instead, the Ly$\alpha$ and CIII]1908 emission lines and a few ISM absorption lines present enough signal to noise (larger than 3) to be identified with certainty. The equivalent widths of the absorption lines in the stacked spectra are also estimated from the ratio between the fluxes integrated within the Gaussian-fit curves and the linearly fitted continuum levels.

\begin{figure}
\centering
\includegraphics[width=100mm]{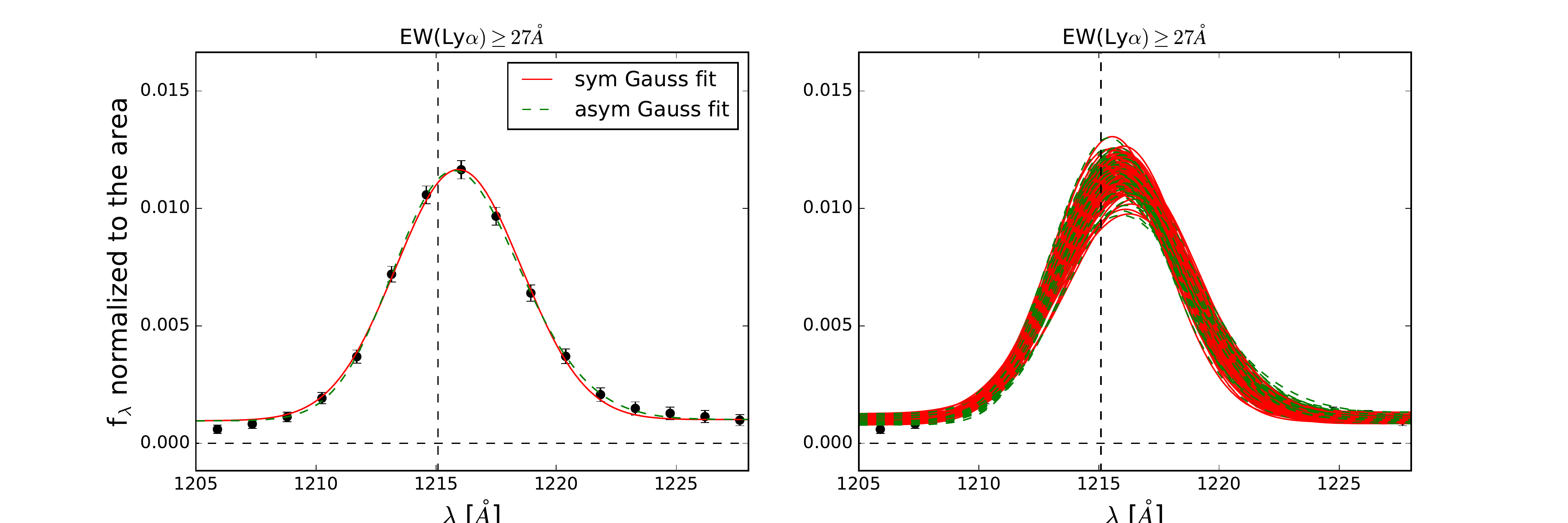}
\includegraphics[width=100mm]{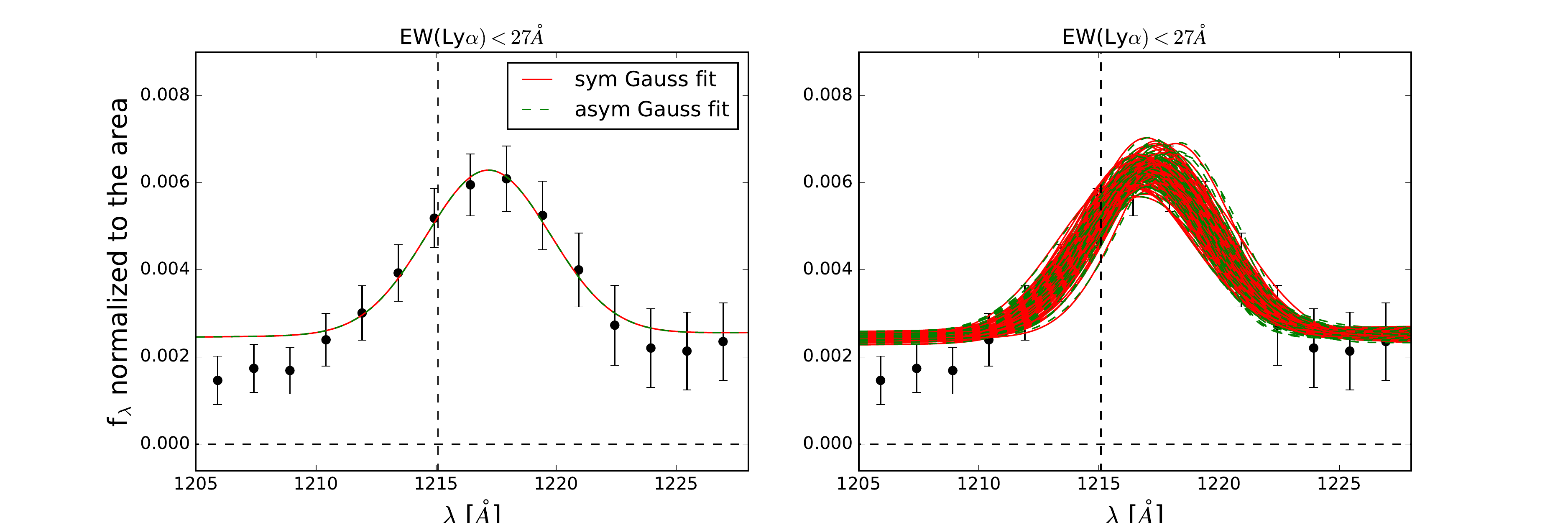}
\caption{Flux density (f$_{\lambda}$) normalized to the area below the curves of the Ly$\alpha$ (black dots) emission line. We present the 1D-stack line of the subsample of the galaxies with EW(Ly$\alpha)\geq27$ {\AA} on the top and with EW(Ly$\alpha)<27$ {\AA} on the bottom panel. In the left panels, we show the symmetric (red) and asymmetric (green) Gaussian best fits. In the right panels, we show the best fits of the 100 realizations of the same line generated to calculate the bootstrap uncertainties. 
For the EW(Ly$\alpha)\geq27$ {\AA} subsample, the Ly$\alpha$ peak shift from the symmetric (asymmetric) Gaussian fit is $196\pm60$ km sec$^{-1}$ ($163\pm63$ km sec$^{-1}$). For the EW(Ly$\alpha)<27$ {\AA} subsample, it is $510\pm98$ km sec$^{-1}$ ($516\pm114$ km sec$^{-1}$) for the symmetric (asymmetric) Gaussian fit.}
\label{fig:CommGauss}
\end{figure}

We also analysed the shape of the Ly$\alpha$ emission line in the stacked 1D and 2D spectra. 
Ly$\alpha$ peaks redward of $z_{\rm sys}$ and we estimated the corresponding velocity shift 
 (Ly$\alpha$ peak shift) 
by fitting a Gaussian curve to the Ly$\alpha$ emission line in the 1D spectrum and by taking the best-fit center of the Gaussian (Fig. \ref{fig:CommGauss}). 

To estimate the errors on the Ly$\alpha$ peak shift, we applied the bootstrap technique. We adopted this technique 
because the uncertainties are calculated directly from the spectra as they are, without the need of choosing an arbitrary method to perturb the spectrum fluxes, as the usual $Monte ~Carlo$ technique would require. Also, it is a random sampling technique that is able to deal with small samples. Following the bootstrap prescription, for the stacked spectrum of each subsample we generated 100 fake stacks, each one constructed from a group of galaxies drawn with a replacement from the subsample that was used for the real stack. On the fake stacks, we performed the same measurements as in the real stack. We estimated the Gaussian-fit parameters in each fake realization and calculated the standard deviation among the 100 realizations.

We explored the possibility of fitting the Ly$\alpha$ emission line with more sophisticated curves, such as a skewed Gaussian function with the following form:
\begin{equation}
f(\lambda) = N e^{ - (\lambda- \overline{\lambda})/2 \sigma^2} \times [1 + erf ((\lambda- \overline{\lambda}) A)]
,\end{equation}
where A is the skewed parameter. However, given the low resolution of our spectra, the best-fit parameters obtained with this skewed Gaussian function are consistent with those obtained with a normal Gaussian fit. An example is shown in Fig. \ref{fig:CommGauss}. We can see that symmetric and asymmetric curve fits produce consistent parameters within the bootstrap errors. Therefore, we perform our analysis of the 1D Ly$\alpha$ profiles with the simplest approach (i.e. the Gaussian fit).

\begin{figure}
\centering
\includegraphics[width=100mm]{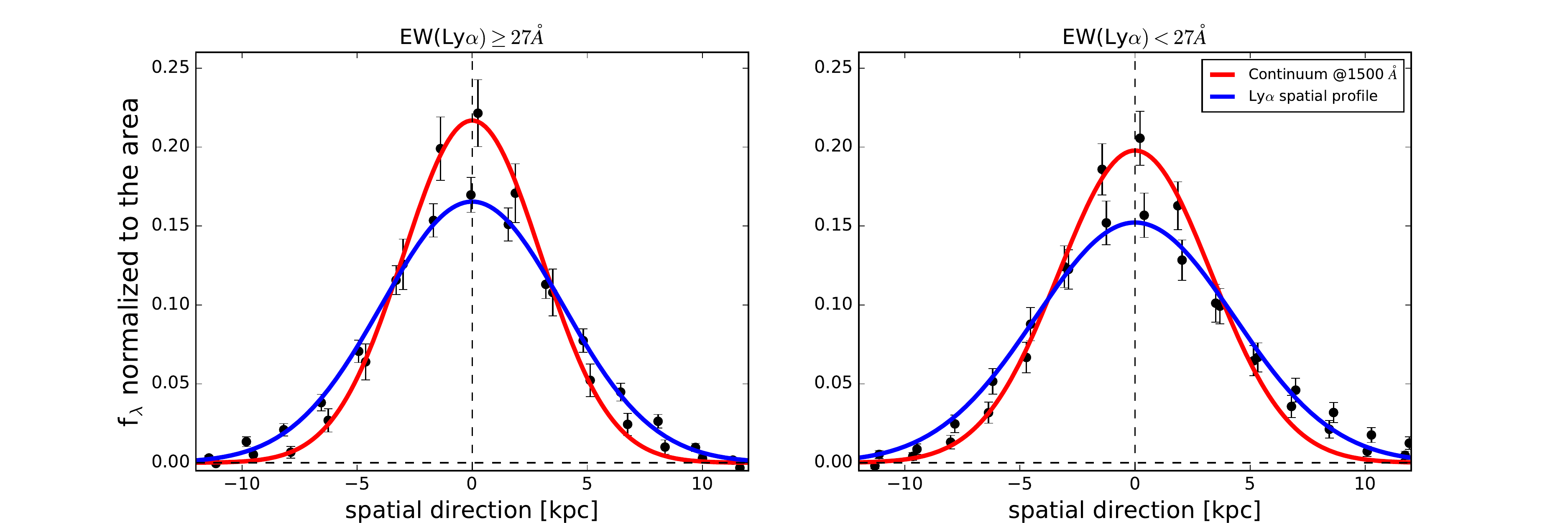}
\caption{Flux density (f$_{\lambda}$) normalized to the area below the curves of the Ly$\alpha$ (black dots and blue curve) and UV continuum (black dots and red curve) spatial profiles. We present the 2D-stack profiles of the subsample of the galaxies with EW(Ly$\alpha)\geq27$ {\AA} on the left and with EW(Ly$\alpha)<27$ {\AA} on the right panel because this is one of the partitions leading to the strongest difference in Ext(Ly$\alpha$ - C). The error bars come from the background in the 2D stack at a spatial-direction position that is either higher or lower than the position of the science spectrum. The continuum is unresolved within the observation point spread function and the Ly$\alpha$ profile is more extended than the continuum profile.}
\label{fig:LyaP}
\end{figure}

%
%
To evaluate the spatial extension of the Ly$\alpha$ emission in each subsample, we first obtained the rest-frame 2D spectra and we aligned them in the spatial direction (Cassata et al., in prep). 
We then averaged the aligned 2D spectra via the IRAF task $imcombine$. The Ly$\alpha$ spatial extension is estimated by fitting a Gaussian curve to the spatial profiles of Ly$\alpha$ and UV continuum around 1500 {\AA} (Fig. \ref{fig:LyaP}). 
We obtain the full width half maximum (FWHM) of the Gaussian best fits (FWHM(Ly$\alpha$) and FWHM(Cont)).
We define the Ly$\alpha$ spatial extension with respect to the continuum as
\begin{equation}
Ext(Ly\alpha-C) = \sqrt{ FWHM(Ly\alpha)^2 - FWHM(Cont)^2} 
\label{eq:diffLyaC}
.\end{equation}

When FWHM(Ly$\alpha) >$ FWHM(Cont), the Ly$\alpha$ spatial profile is more extended than the rest-frame UV continuum and Ext(Ly$\alpha$-C) is defined. We adopt this definition to implicitly consider the deconvolution with the point spread function (PSF) of the observations, which is difficult to quantify in the stacked spectrum. In fact, a subtraction in quadrature of the PSF FWHM from FWHM(Ly$\alpha$) and FWHM(Cont) cancels out in our 
 formula\footnote{The Ext(Ly$\alpha$-C) we measure could be a lower limit of the true Ly$\alpha$ spatial extension. In fact, VUDS spectra are corrected for slit losses based on continuum magnitudes \citep{LeFevre2015}, but additional losses could affect Ly$\alpha$ if its emission is significantly more extended than the continuum \citep[e.g.][]{Huang2016}. 
For instance, Ly$\alpha$ emission elongated 
perpendicularly to the spatial direction 
will be strongly affected by slit losses. 
Since we are considering stacks it is reasonable that severe losses will be compensated.
}

This definition does not consider the difference in surface brightness of galaxies at different redshifts. This could be a limit of our approach. However, we emphasize that the median $R$ magnitude and $z_{\rm sys}$ are very similar for all the subsets. 
 
In Table \ref{tab:sub}, we show the properties of each subsample, such as the median systemic redshift, $R$ magnitude, SFR, star formation rate surface density ($\Sigma_{\mathrm {SFR}}$), stellar mass, and SiII velocity offset ($\Delta$v), equivalent width of the SiII1526 absorption line, CIII]1908, Ly$\alpha$ equivalent widths, Ly$\alpha$ spatial extension, and Ly$\alpha$ peak shift of the stacks. Star formation rate and stellar mass are obtained from the SED fitting; 
the star formation rate surface density \citep{Heckman2015}  is obtained from SFR and the PSF-corrected half-light-radius,
\begin{equation}
\Sigma_{\mathrm {SFR}} [M_{\odot} yr^{-1} kpc^{-2}] = \frac{SFR}{2\pi r_{\mathrm {1/2}}^2}, 
\end{equation}
where 
r$_{\mathrm {1/2}}$ is the PSF-corrected half-light radius of a galaxy as defined and estimated in \citet{Ribeiro2016}. 

The large error bar on the median stellar masses in the table indicate that the galaxies in each subset can be characterized by masses in the entire range of stellar masses, log(M$_{*}$/M$_{\odot}$)=[8.5-11]. 
\subsection{Kinematics features}
\label{sec:kin}

To quantify the ISM kinematics, 
we focused on the velocity offset between the central wavelength of the LIS lines and the systemic redshift of the stacked spectra. We chose the two absorption lines with the highest signal to noise in the majority of our stacked spectra, SiII1260 and SiII1526 (see Fig. \ref{stackedspectra1} and \ref{stackedspectra2}). 

We first set the systemic rest frame of the stacked spectrum in the following way. 
We fit the stacked-spectrum CIII]1908 line by assuming that the ratio between the fluxes of the lines of the doublet 
is 1.5 \citep[Sec. \ref{sec:Data}][]{Keenan1992}\footnote{An uncertainty in the flux ratio (F(1907)/F(1909)) of the CIII doublet could be translated into an uncertainty in the systemic redshift. As calculated in \citet{Keenan1992}, F(1907)/F(1909) varies from 1.5 to 0 for $1.5<$ log(n$_{e}) <6.5$ cm$^{-3}$ and it is unchanged for $5000<$ T$_{e} <20000$ K (their Fig. 1). For log(n$_{e}) <3.5$ cm$^{-3}$, the ratio is constant as F(1907)/F(1909)=1.5. \citet{Sanders2016} measured the electron density of a sample of $z\sim2.3$ star-forming galaxies with SFRs that are consistent with those of our sample. By using the flux ratios of nebular doublets, they infer a typical n$_{e}$ of $\sim$250 cm$^{-3}$ (their Fig. 3 and section 3.4). Previous measurements had calculated values up to 1000 cm$^{-3}$, for which we can still adopt F(1907)/F(1909)=1.5 \citep{Talia2012} and infer a reliable $z_{\rm sys}$.}. 
We then fit negative Gaussian$+$continuum functions to the SiII1260 and SiII1526 systems. 
%
We assumed that the two SiII lines have the same velocity because they are transitions of the same element
, but a different amplitude, and we searched for the best-fit velocity. We adopted the noise of the stacked spectrum in the minimization procedure.

The velocity offset between the LIS and the systemic redshift is calculated from the best-fit central wavelength of the SiII Gaussian fit, $l0_{best}(SiII)$,
\begin{equation}
\Delta v(SiII)=c \times \frac{l0_{\mathrm{best}}(SiII)-l\_SiII_{\mathrm{air}}\_z_{\mathrm{sys}}}{l\_SiII_{\mathrm{air}}\_z_{\mathrm{sys}}},
\label{eq:deltav}
\end{equation}

where $l$\_$SiII_{air}$\_$z_{\rm sys}$ is the theoretical wavelength of SiII in the air and $c$ is the speed of light in km sec$^{-1}$.
There can be a significant negative value of $\Delta$v(SiII). This is interpreted as the fact that the neutral gas is predominantly flowing out of the galaxy; there could be a gas component either at systemic redshift or inflowing, which is hidden within the outflowing gas.
A $\Delta$v(SiII) value consistent with 0 km sec$^{-1}$ can indicate either a static medium or a medium in which there are equal inflowing and outflowing gas components.

To estimate the errors on $\Delta$v, as well as on the equivalent widths, and Ext(Ly$\alpha$-C), we also applied the bootstrap technique. 
%
As for the Ly$\alpha$ peak shift, we estimated the Gaussian-fit parameters in each fake realization and calculated the standard deviation among the 100 realizations. The standard deviation of $l0_{best}$($SiII$) provides the error on $\Delta$v. 

\section{Results}
\label{sec:Results}

%
We present here the measurements obtained applying the methods described in section \ref{sec:Method}.
The measured quantities involved the 1D, such as equivalent widths, gas velocities, and Ly$\alpha$ peak shift, and 2D, such as Ext(Ly$\alpha$-C), stacks. 
The scope of these measurements is to study simultaneously the Ly$\alpha$ spectral and spatial escape from the typical galaxy in each subset listed in Table \ref{tab:sub}. 
%

\begin{figure*}
\includegraphics[width=210mm]{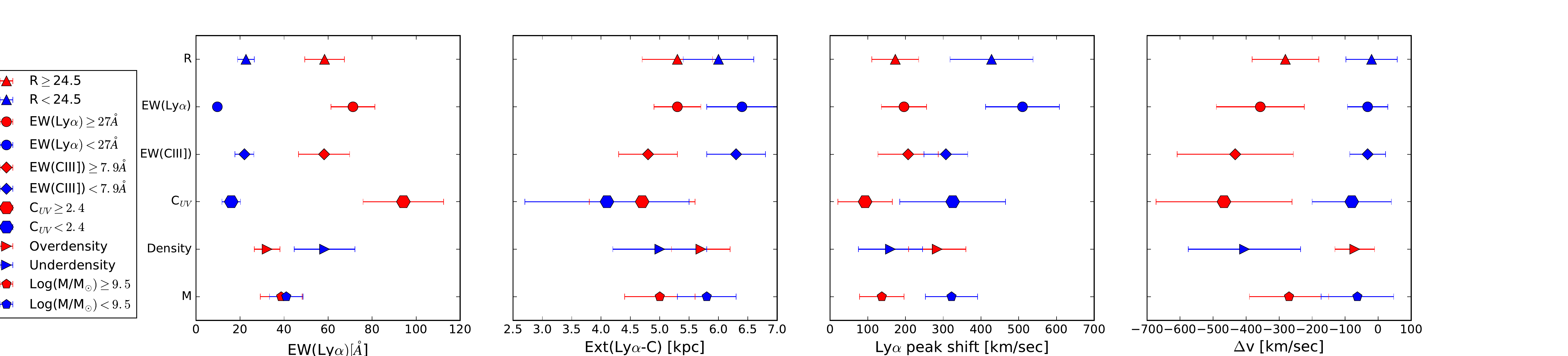}
\caption{Measurements of EW(Ly$\alpha$) ($first~ panel$), Ext(Ly$\alpha$-C) ($second~ panel$), Ly$\alpha$ peak shift ($third~ panel$), and $\Delta$v ($fourth~ panel$), together with the bootstrapping (see text) error bars for the subsamples listed in Table \ref{tab:sub}. The symbols correspond to the subsets listed in Table \ref{tab:sub}: 
galaxies fainter(brighter) than the median R magnitude of the entire sample are shown as red(blue) triangles; galaxies characterized by EW(Ly$\alpha$) larger(smaller) than the median value are shown as red(blue) filled circles; galaxies characterized by EW(CIII]1908) larger(smaller) than the median value as red(blue) diamonds; galaxies with rest-frame UV concentration larger(smaller) than the median value as red(blue) hexagons; galaxies with Ext(Ly$\alpha$-C) larger(smaller) than the median value as red(blue) squares; galaxies in overdense(underdense) regions as red(blue) tilted triangles; and galaxies more(less) massive than the median value as red(blue) pentagons.
}
\label{fig:Paramsig}
\end{figure*}

Fig. \ref{fig:Paramsig} shows the significance of the differences of the Ly$\alpha$ parameters and $\Delta$v within the pairs of subsamples. The subsamples of the galaxies that are faint and concentrated in the rest-frame UV characterized by 
EW(CIII]1908) $\geq7.9$ {\AA} have EW(Ly$\alpha$) more than 2$\sigma$ higher than the paired subsamples. The stacks of the galaxies located in overdense and underdense regions, and those of the most and the least massive galaxies show consistent EW(Ly$\alpha$) values within the uncertainties. 

Differences of the order of 2$\sigma$ are observed for $\Delta$v, except for the pairs of subsets separated based on the median value of the stellar mass and 
the rest-frame UV concentration for which the velocity offsets are consistent. 
The galaxies brighter than $R=24.5$ and more extended than the median in the rest-frame UV, which are characterized by EW(Ly$\alpha)<27$ {\AA}, EW(CIII]1908) $<7.9$ {\AA}, and localized in overdense regions, have a medium that is consistent with being static. The paired subsamples show a medium with a net flow out of the galaxy with a velocity larger than 300 km sec$^{-1}$. 

In terms of Ext(Ly$\alpha$-C) and Ly$\alpha$ peak shift, differences within pairs can be observed, but tend to be less significant for some subsets.
One of the limiting factors is that the spectra are obtained with VIMOS in low-resolution mode with a sampling of about 1 {\AA}/pixel ($\sim$250 km sec$^{-1}$ at $\lambda$Ly$\alpha$ and $\sim$150 km sec$^{-1}$ at $\lambda$CIII]1908). 
In addition, the sky subtraction technique \citep[a low-order spline fit along the slit for each wavelength sampled;][]{LeFevre2015}  could produce an over subtraction near strong lines, such as a strong Ly$\alpha$.
The over subtraction (seen as negative pixel values) is particularly evident in stacked spectra and it can limit the accuracy of the spatial extent measurement.
%
The difference of Ext(Ly$\alpha$-C) we measure is generally of the order of 1$\sigma$. For the subsets of the most/least UV concentrated galaxies, the parameters are consistent. 
The stacks of the galaxies that are bright in the rest-frame UV, faint in Ly$\alpha$, and less massive than the median are characterized by Ly$\alpha$ peak shifts 2$\sigma$ larger than the UV faint, Ly$\alpha$ bright, and more massive galaxies.

In general, the error bars of all the measurements in the stacks of the most/least UV concentrated galaxies are large. One of the reasons is the smaller number of sources in comparison to the other subsamples. 

\begin{figure*}
\centering
\includegraphics[width=150mm]{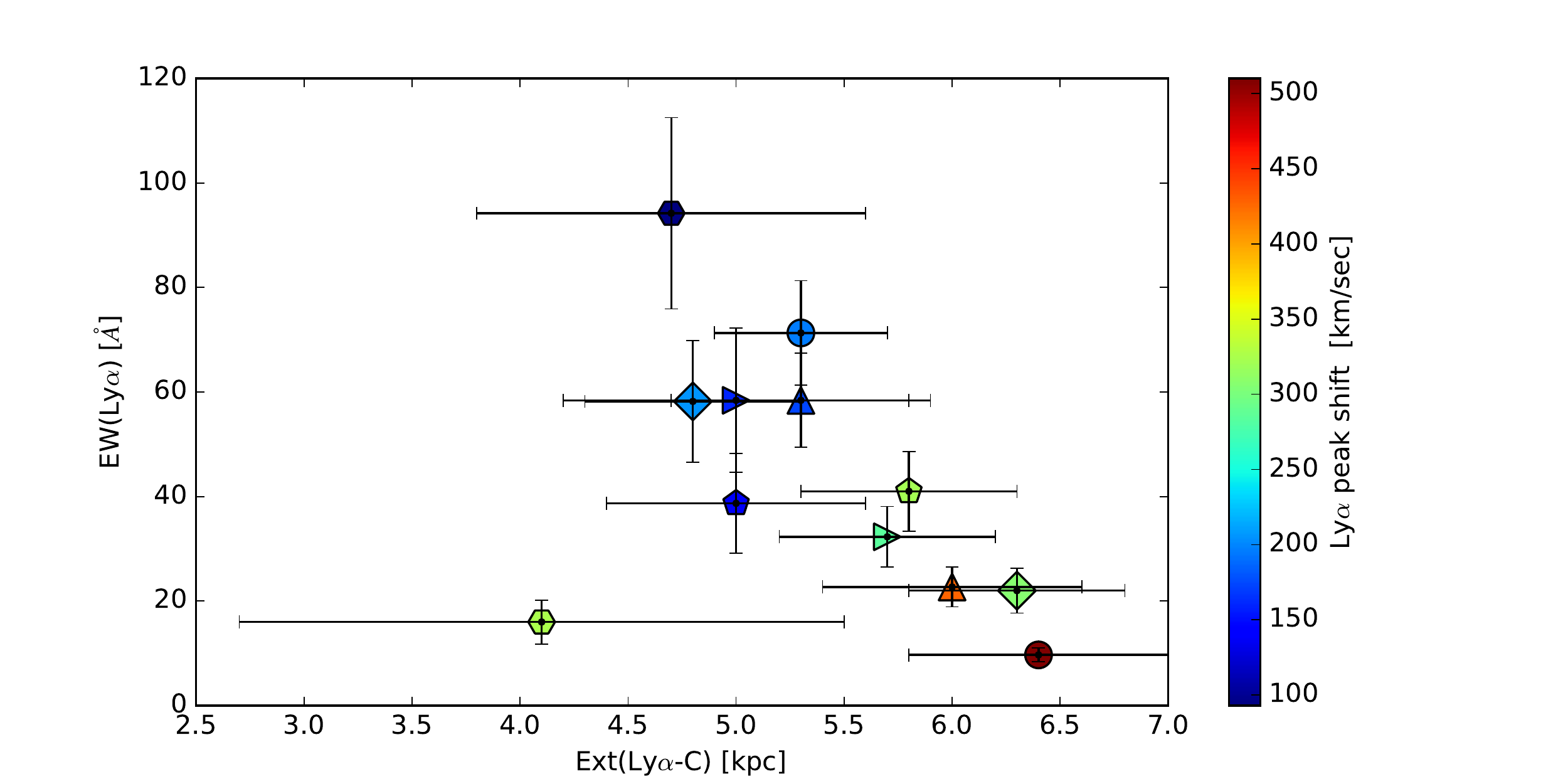}
\caption{Ly$\alpha$ equivalent width vs. Ly$\alpha$ spatial extension in kpc. The colour coding follows the values of the Ly$\alpha$ peak shift. The symbols represent the different subset listed in Table \ref{tab:sub} as shown in Fig. \ref{fig:Paramsig}.
}
\label{fig:EWExtLyapeak}
\end{figure*}

Fig. \ref{fig:EWExtLyapeak} is the most significant figure in the paper; this figure shows the three Ly$\alpha$ parameters that are outputs of the phenomena that allow the propagation inside and escape out of a galaxy. For the subsets for which the differences in the three parameters are significant (see Fig. \ref{fig:Paramsig}), EW(Ly$\alpha$) anti-correlates with Ext(Ly$\alpha$-C) and Ly$\alpha$ peak shift. This trend is discussed in the following section. 
The bootstrap error bars of Ext(Ly$\alpha$-C) for the galaxies with C$_{UV}<2.4$ are the largest. 
Also, the galaxies in the C$_{UV}<2.4$ subset are characterized by a larger variety of Ly$\alpha$ profiles and EW(Ly$\alpha$) than the C$_{UV}\geq2.4$ subset. The median of the EW(Ly$\alpha$) measured in the individual spectra of the galaxies belonging to the C$_{UV}<2.4$ subset is $13\pm6$ {\AA}, while for the individual spectra of the galaxies in the C$_{UV}\geq2.4$ subset is $83\pm21$ {\AA}.

\begin{figure*}
\centering
\includegraphics[width=200mm]{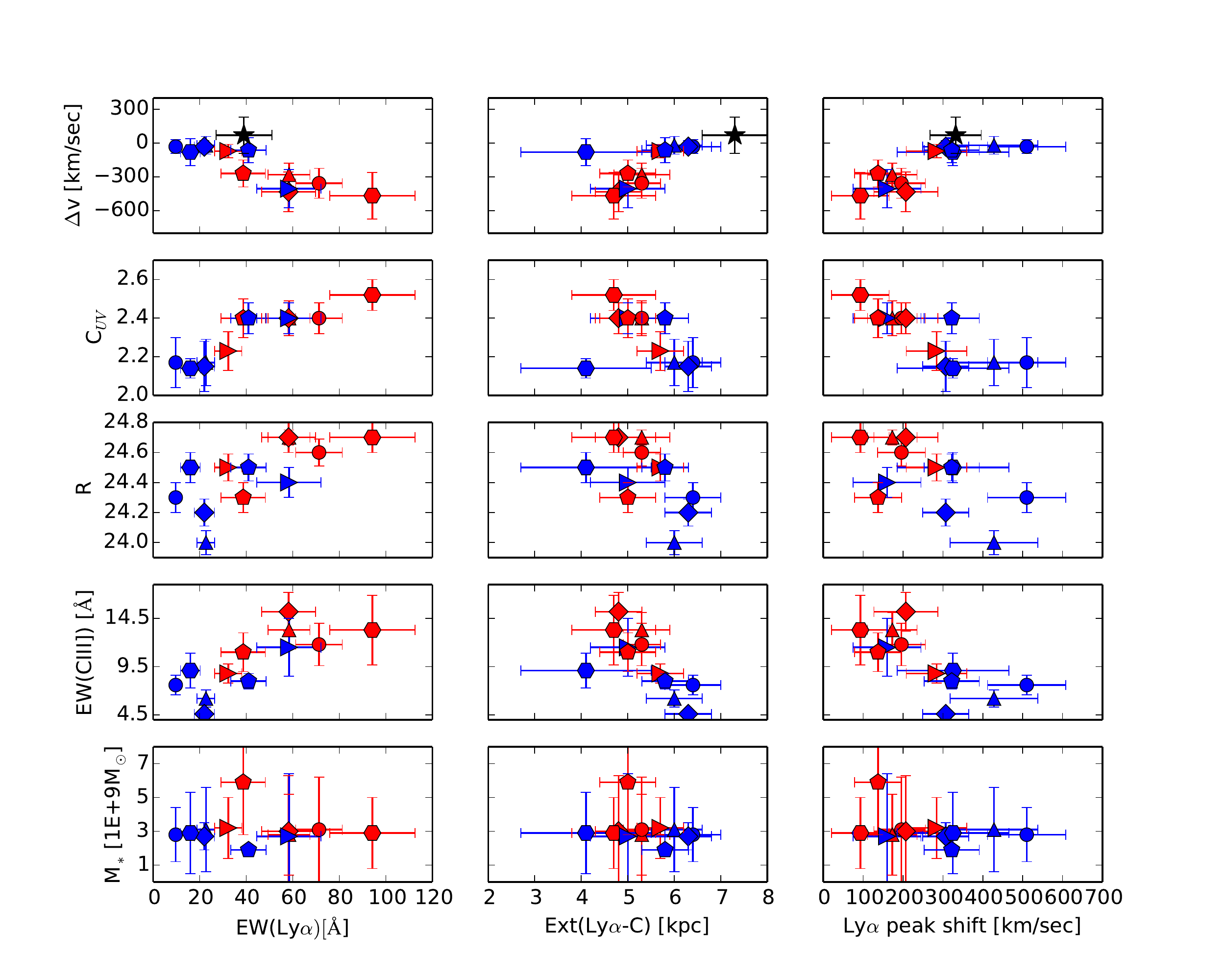} 
\caption{Stack physical properties vs.  Ly$\alpha$ properties. From the top row: The plots show $\Delta$v, rest-frame UV concentration, rest-frame UV magnitude, EW(CIII]1908), and stellar mass vs. EW(Ly$\alpha$) ($first~ column$), Ext(Ly$\alpha$-C) ($second~ column$), and Ly$\alpha$ peak shift ($third~ column$). Symbols are as in Fig. \ref{fig:Paramsig}. 
For each subset, $\Delta$v, EW(Ly$\alpha$), EW(CIII]1908), and Ly$\alpha$ peak shift are measured in the stacks, while C$_{UV}$, R, and M$_{*}$ correspond to the median values of the galaxies within the subset. 
The black star represents a subset of sources with extreme Ext(Ly$\alpha$-C) in the individual spectra. These sources are characterized by moderate EW(Ly$\alpha$), Ly$\alpha$ peak shift larger than 300 km sec$^{-1}$ and $\Delta$v consistent with zero.
}
\label{fig:parprop}
\end{figure*}

In Fig.\ref{fig:parprop}, we wish to enlighten any possible trend between physical and Ly$\alpha$ properties. 
The first row of the figure shows that large outflow velocities seem to favour large values of EW(Ly$\alpha$), small values of Ext(Ly$\alpha$-C) and small  Ly$\alpha$ peak shifts. 
In the following section, we discuss the role of outflows in the generation of a Ly$\alpha$ emission with certain values of equivalent width, spatial extension, and peak shift.
Galaxies that are very concentrated and faint in the rest-frame UV 
%
%
have the largest outflow velocities ($\Delta$v $\geq$ 300 km sec$^{-1}$) and EW(Ly$\alpha)\geq 40$ {\AA}. 
 The Ly$\alpha$ photons emitted from galaxies that are bright in the rest-frame UV show larger Ext(Ly$\alpha$-C) values and larger Ly$\alpha$ peak shifts than the paired subsets. 
Low-EW(CIII]1908) sources are characterized by a large HI column density and hence 
large Ext(Ly$\alpha$-C) values. 
These sources also show larger Ly$\alpha$ peak shifts than the subset of high-EW(CIII]1908) galaxies. 
We do not observe significant trends between Ly$\alpha$ properties and stellar mass. 
%

\section{Discussion}
\label{sec:Discussion}

The aim of this work is to investigate the effect of HI gas kinematics and other gas/galaxy properties, such as the HI column density, on the escape and large-scale propagation of Ly$\alpha$ photons in star-forming galaxies. We investigate the gas kinematics in terms of the velocity of the outflowing HI gas and we study the dependency of the outflow velocity inferred by SiII as a function of the equivalent width, spatial extension, and red peak shift of Ly$\alpha$  with respect to the systemic redshift. 

Before interpreting the kinematics of the gas in the galaxies of our sample, we comment on the dependency of outflow velocity on star formation properties (Sec. \ref{sec:SF}). Then, we propose a model to explain our results (Sec. \ref{sec:model}) and we interpret the observed trends in Sec. \ref{sec:interpret}.

\subsection{Star formation properties}
\label{sec:SF}

%
Based on a simple model, when the pressure created by supernovae, stellar winds, and the radiation field from the star formation exceed the weight of the galaxy disk, an outflow can be launched perpendicular to the galactic plane. The strength of this outflow depends on the SFR per unit of area. 
\citet{Heckman2001} proposed the existence of a $\Sigma_{\mathrm {SFR}}$ threshold above which a galaxy would be able to support outflows. 
Later on, they investigated the scaling relations of the outflow velocity with physical parameters and found that it weakly correlates with stellar mass, but strongly with SFR and $\Sigma_{\mathrm {SFR}}$ \citep{Heckman2015, Alexandroff2015, Heckman2016}. 
\citet{Newman2012} studied the kinematics in a sample of $z\sim2$ star-forming galaxies with SFRs from a few tens to a few hundreds of M$_{\odot}$ yr$^{-1}$ and M$_*$ of the order of $2-100 \times 10^{9}$ M$_{\odot}$.
They found that the strongest outflows are observed for massive (M$_*> 10^{10}$ M$_{\odot}$), high-SFR (SFR $>100$ M$_{\odot}$yr$^{-1}$), compact (r$_{1/2}<3$ kpc), face-on galaxies, which, by consequence, are characterized by $\Sigma_{\mathrm {SFR}} >1$ M$_{\odot}$ yr$^{-1}$ kpc$^{-2}$. 
\citet{Chisholm2015} showed that the strongest outflows can also arise from merging events.

In Table \ref{tab:sub}, we report the median values of SFR and SFR per unit of area, 
obtained from the same definition 
in \citet{Heckman2015} . 
We do not see a clear proportionality between $\Sigma_{\mathrm {SFR}}$ and outflow velocity. However, we can see that in the subsamples for which the difference between the $\Delta$v values are more significant, there are also notable differences in the $\Sigma_{\mathrm {SFR}}$ values. In particular a higher value of $\Delta$v is related to a higher value of $\Sigma_{\mathrm {SFR}}$ (first three rows of Table \ref{tab:sub}). 
%


The SFR and size conditions of the typical galaxy of our subsamples could allow the formation of outflows and we want to investigate how much they contribute to the escape and propagation of Ly$\alpha$ photons.

\subsection{Theoretical model}
\label{sec:model}

We compare our observational results with the theoretical model by \citet{V2006}. We chose this model because it allows us to predict the quantities we are considering in this work, such as Ly$\alpha$ equivalent width, spatial extension, and peak shift, with respect to HI column density and outflow velocity.  

The model predicts the shape of the Ly$\alpha$ emission line, by assuming an expanding, spherical, homogeneous, and isothermal shell of neutral hydrogen surrounding a central starburst. The stars emit UV continuum and the Ly$\alpha$ radiation is produced in the ISM. The expanding shell is characterized by an expansion velocity, V$_{\mathrm exp}$, which can mimic the outflow velocity. By following a $Monte ~Carlo$ approach, Ly$\alpha$ photons follow a random path encountering HI atoms and dust grains. The model produces the Ly$\alpha$ spectrum out of the galaxy depending on the shell parameters, such as V$_{\mathrm exp}$,  HI column density, and dust absorption optical depth. 

For increasing HI column density, the Ly$\alpha$ main peak is increasingly red-shifted \citep{Verhamme2015}.
 However, outflow velocities larger than 300 km sec$^{-1}$ could allow some of the Ly$\alpha$ photons to escape at the systemic redshift 
\citep[the Ly$\alpha$ peak would be significantly red-shifted if N$_{\mathrm {HI}}$ $>10^{20}$ cm$^{-2}$; see Fig. 2 in][]{Verhamme2015}.
An HI gas characterized by those high velocities and low to moderate N$_{\mathrm {HI}}$ is not efficient in scattering Ly$\alpha$ photons. As a consequence, the majority of the Ly$\alpha$ flux is peaked in the center of the galaxy and does not produce an extended Ly$\alpha$ spatial profile. The models from \citet{BH2010} also indicate that the Ly$\alpha$ emission becomes more centrally peaked with the increase of the velocity of the HI bulk \citep[see also][]{Gronke2016}. 

With the \citet{V2006} model, it is possible to study Ly$\alpha$ spatial extension and equivalent width at given shell parameters. The model predicts that the Ly$\alpha$ spatial extension is proportional to N$_{\mathrm {HI}}$ and depends on the scattering capability of the HI gas. For N$_{\mathrm {HI}}$ $>10^{19}$ cm$^{-2}$, the scattering is efficient. So, a very large V$_{\mathrm exp}$ would be needed to make a compact Ly$\alpha$ emission. For N$_{\mathrm {HI}}$ $<10^{19}$ cm$^{-2}$, the Ly$\alpha$ emission is spatially compact independently on V$_{\mathrm exp}$ (Verhamme et al. in prep). 
%
Also, it is possible to predict that EW(Ly$\alpha$) increases when the optical depth of the ISM decreases. 
Since dust completely absorbs Ly$\alpha$ photons, it reduces the amount of photons coming out towards the center and those that are distributed on a large scale. As a consequence EW(Ly$\alpha$) and Ext(Ly$\alpha$-C) decreases. On the other hand, the HI gas scatters Ly$\alpha$ photons away from the line of sight (Verhamme et al. in prep). 

%

\subsection{Interpretation of the observed small outflow velocity, small Ly$\alpha$ equivalent width, and large Ly$\alpha$ spatial extension}
\label{sec:interpret}

%
As described in Sec. \ref{sec:Results}, 
the subsamples of the galaxies that are bright in UV and faint in Ly$\alpha$ show Ly$\alpha$ peak shifts up to 500 km sec$^{-1}$, Ext(Ly$\alpha$-C) larger than 6 kpc, EW(Ly$\alpha)<40$ {\AA}, and are characterized by $\Delta$v that is consistent with a static medium. Instead, the subsamples of $R$ faint and Ly$\alpha$ bright galaxies, characterized by large $\Delta$v, and therefore large outflow velocities, present the Ly$\alpha$ red peak close to the systemic redshift, less extended Ly$\alpha$ spatial profiles, and larger Ly$\alpha$ equivalent widths. These behaviours are also seen for the subsamples that are separated based on rest-frame UV concentration and local density. In particular, for the most compact galaxies we measured $\Delta$v consistent with an outflowing gas and larger EW(Ly$\alpha$) than for the least compact. The galaxies located in underdense regions present larger $\Delta$v, Ly$\alpha$ peak shift, Ext(Ly$\alpha$-C), and smaller EW(Ly$\alpha$) than the galaxies located in overdense regions. 
For a subset of 10 galaxies with extreme Ext(Ly$\alpha$-C) values (black stars in Fig.\ref{fig:parprop}), we measure a $\Delta$v value consistent with a static medium and EW(Ly$\alpha)=40$ {\AA}. 

\begin{figure*}
\centering
\includegraphics[width=60mm]{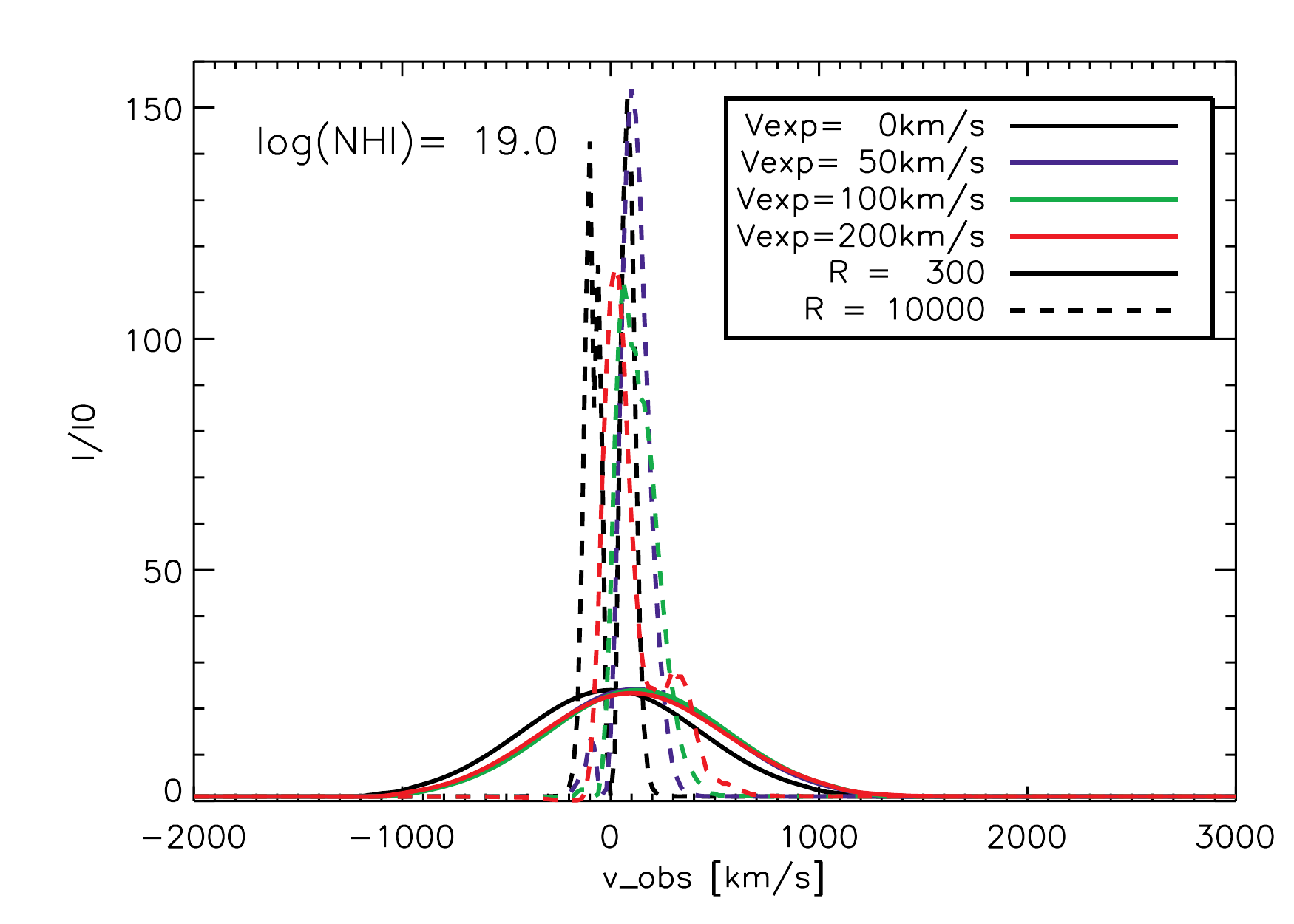}
\includegraphics[width=60mm]{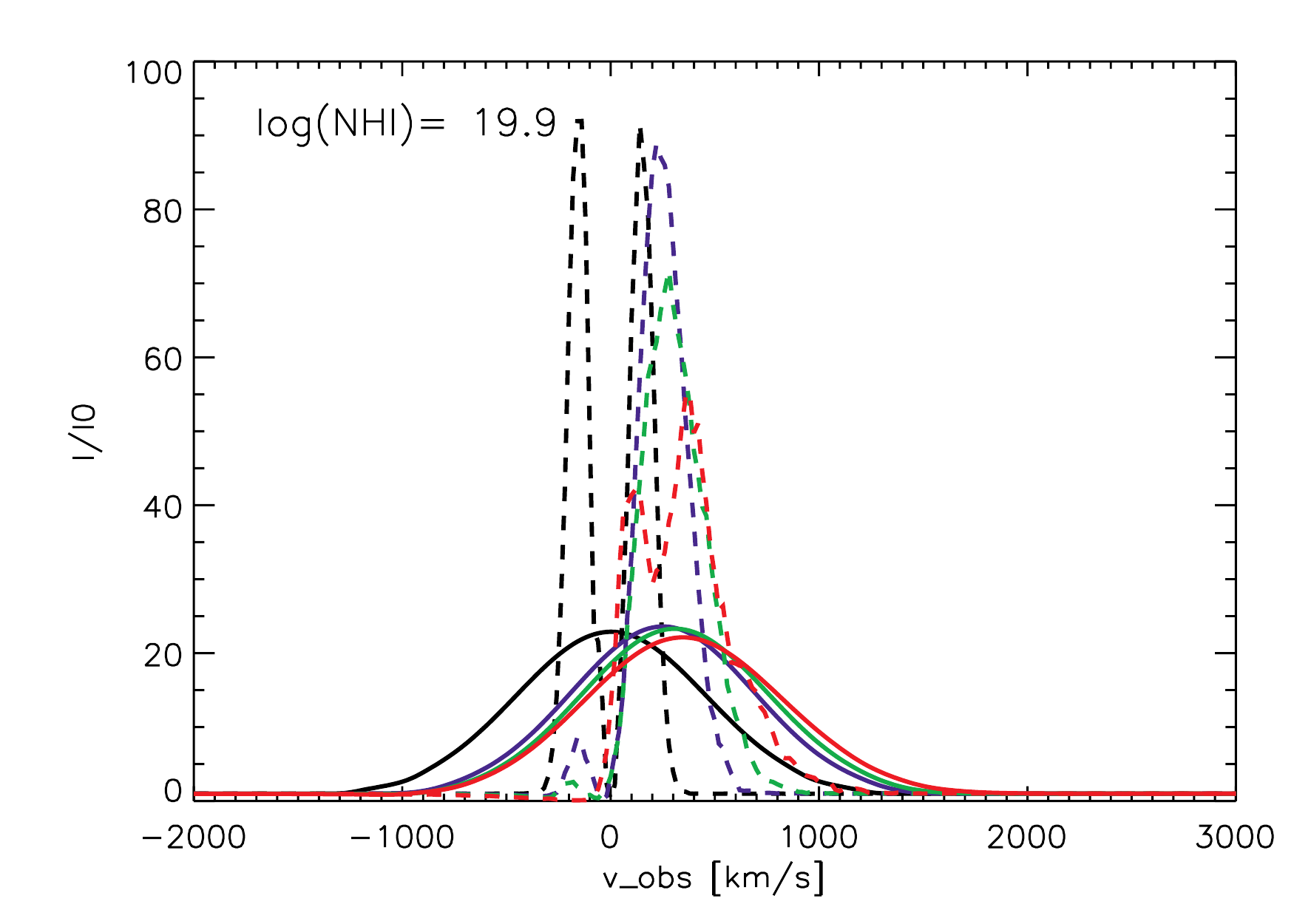}
\includegraphics[width=60mm]{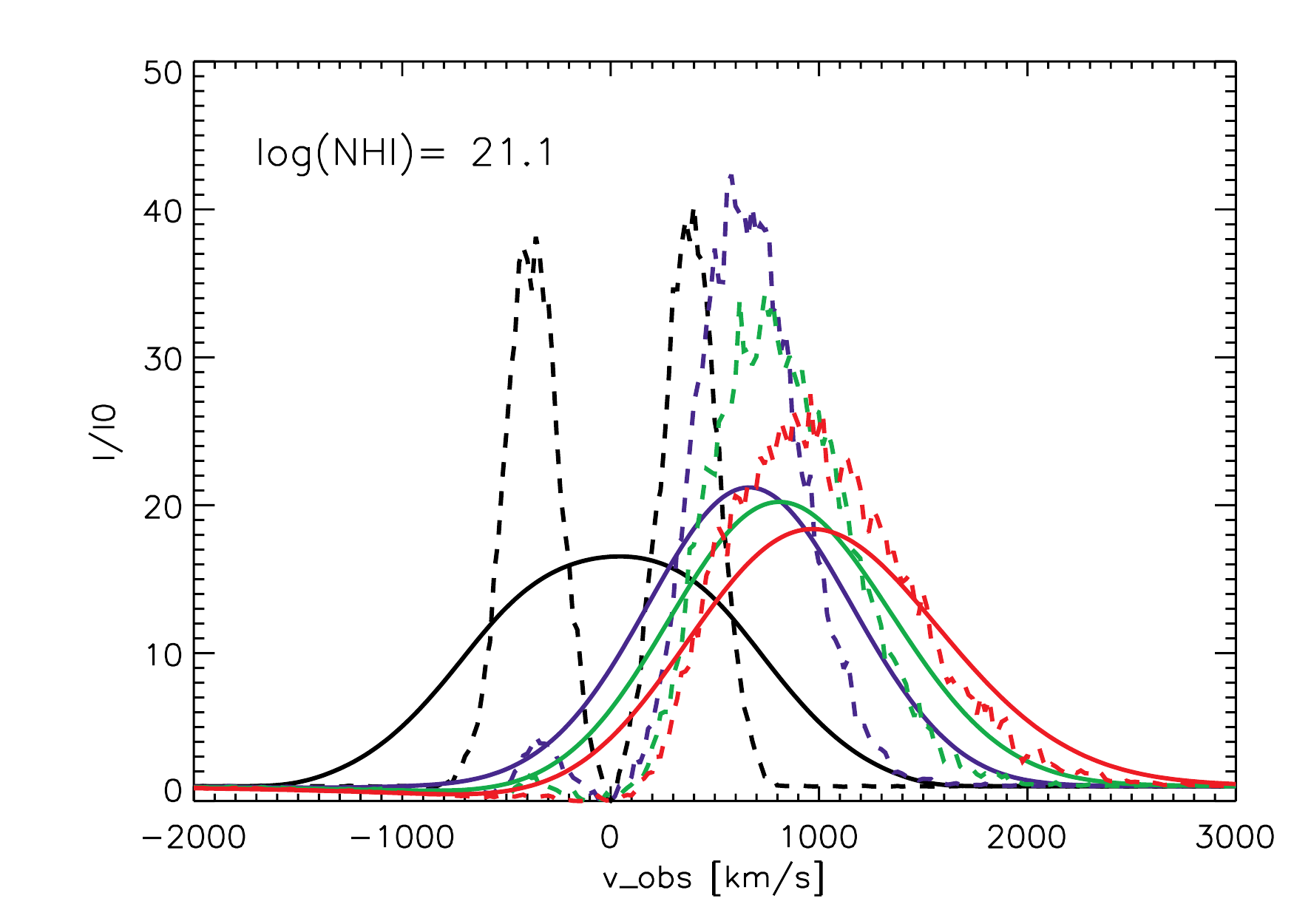}
\includegraphics[width=60mm]{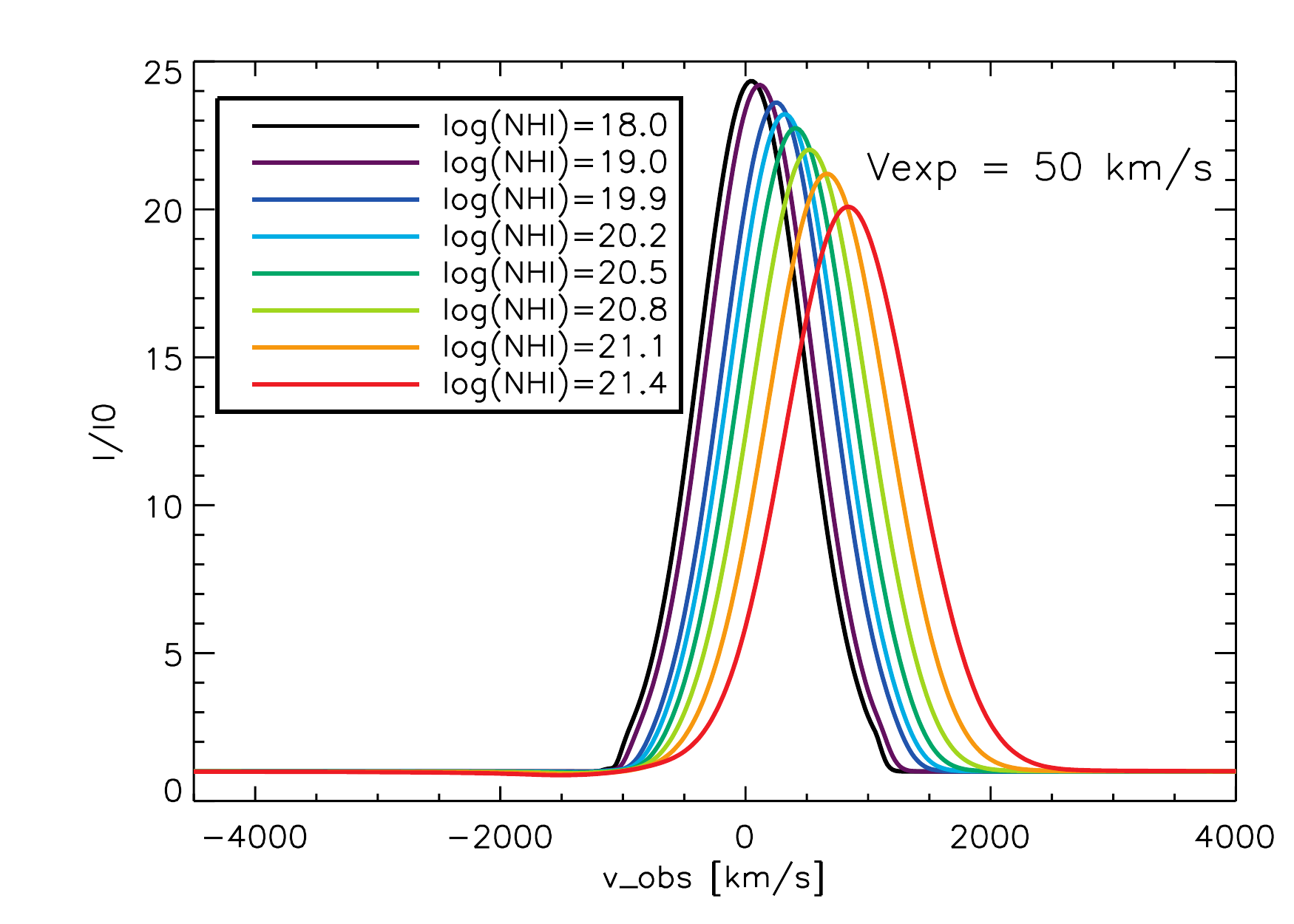}
\includegraphics[width=60mm]{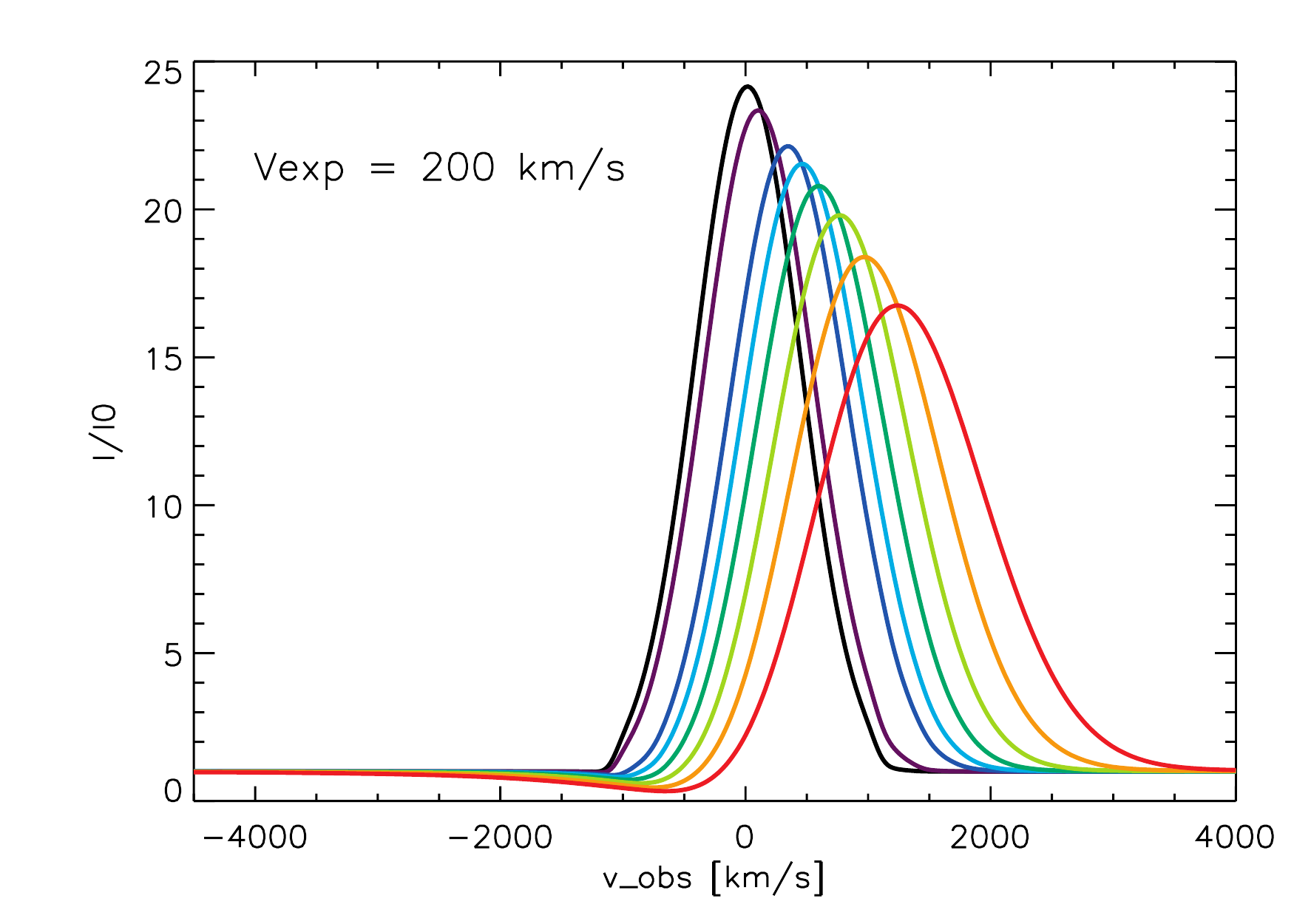}
\includegraphics[width=60mm]{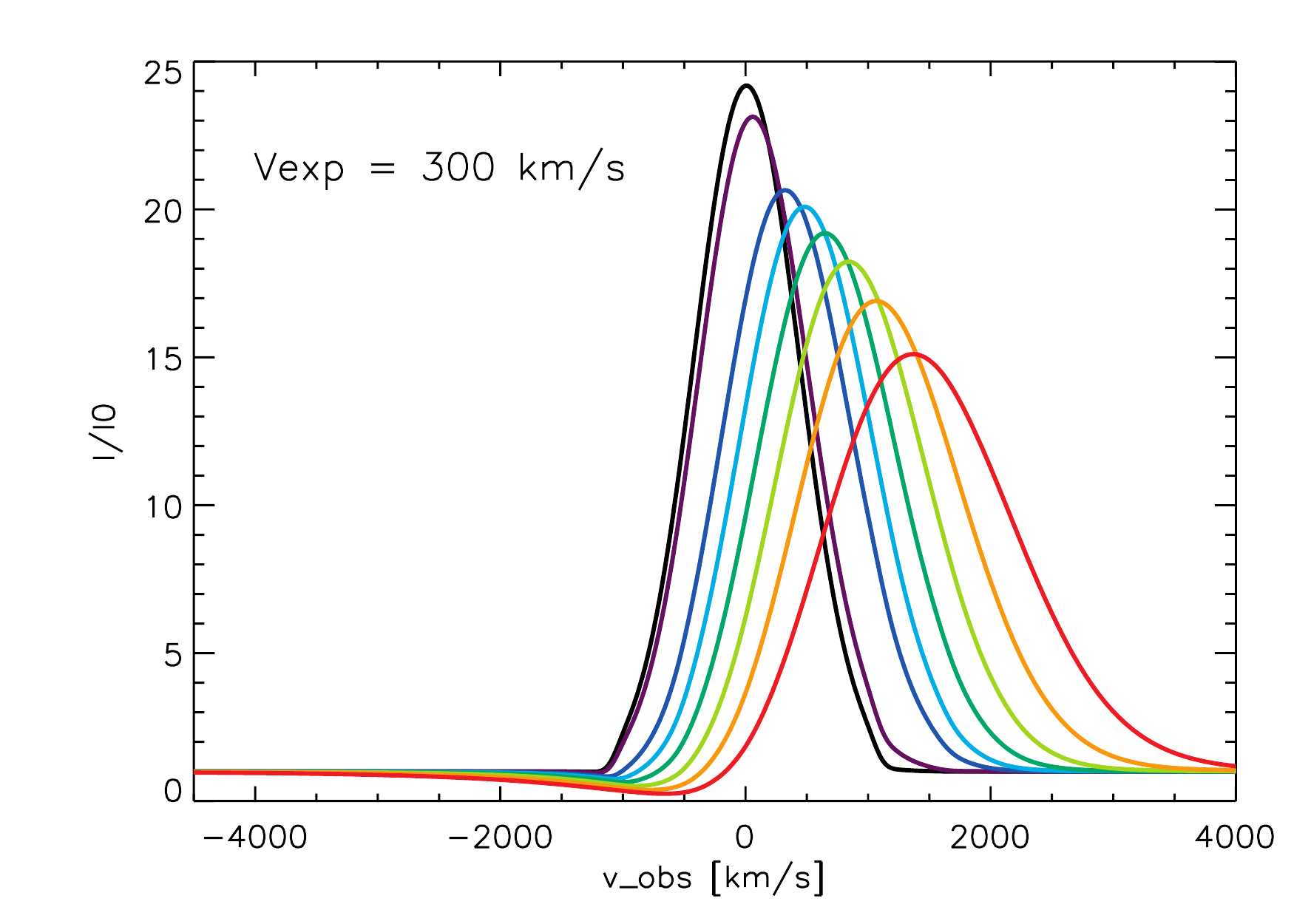}
\caption{Theoretical 1D Ly$\alpha$ emission line (in velocity space) profiles obtained with the adopted (dust-free) shell model. The Ly$\alpha$ profiles (I) are normalized to the level of the continuum (I0). The maximum resolution profiles are shown as dashed lines. Solid lines correspond to VUDS resolution, R $\sim300$. $Upper~ row:$ we assumed different shell expansion velocities and three HI column densities, log(N$_{\mathrm {HI}}$)=19 ($left~panel$), 20 ($middle~ panel$), and 21 ($right~panel$). 
$Lower~row:$ we assumed different values of HI column density and three shell expansion velocities, V$_{\mathrm exp}$ = 50 ($left~panel$), 200 ($middle~ panel$), and 300 ($right~panel$) km sec$^{-1}$. 
}
\label{fig:AnnevsVNHI}
\end{figure*}

From the first row of Fig. \ref{fig:parprop}, we can see that kinematics is important for the propagation and escape of Ly$\alpha$ photons. 
However, to investigate if the kinematics alone is enough to explain the trend of large Ly$\alpha$ peak shift, large Ext(Ly$\alpha$-C), and small EW(Ly$\alpha$), 
we compare our results with the theoretical models by \citet{V2006} (Sec. \ref{sec:model}). We first simulate the 1D Ly$\alpha$ emission line for different shell expansion velocities and for three HI column densities (Fig. \ref{fig:AnnevsVNHI} upper row). Expansion velocities of 50-400 km sec$^{-1}$ would produce near-overlapping Ly$\alpha$ peaks if the spectral resolution were comparable to that of VUDS (R $\leq300$) and N$_{\mathrm {HI}} \leq10^{19}$ cm$^{-2}$. 
This tells us that the kinematics alone would have an effect that is too small on the Ly$\alpha$ peak shifts to be seen at VUDS resolution. 
For small HI column densities, shift differences of the order of 100 km sec$^{-1}$ are not discernible. Increasing N$_{\mathrm {HI}}$, the separation between Ly$\alpha$ peaks at different V$_{\mathrm {exp}}$ enlarges. In this model, the effect of the intergalactic medium (IGM) on the shape of the Ly$\alpha$ line is not included. It has been shown that the effect of the IGM is marginal in comparison to that of the ISM, even more if the ISM gas is already outflowing \citep[e.g.][]{Dijkstra2016}. In fact, a typical red shift of the 
Ly$\alpha$ line is also seen in star-forming galaxies in the local Universe \citep[Orlitova et al. in prep;][for already published examples]{Henry2015, Verhamme2017, YangMalhotra2017}. The correlation between Ext(Ly$\alpha$-C) and Ly$\alpha$ peak shift also tells us that Ly$\alpha$ photons are produced within the galaxy. 

Then, we simulate the same line by varying the HI column density (Fig. \ref{fig:AnnevsVNHI}, lower row). The Ly$\alpha$ peak shift is much larger than the expansion velocity. For velocities of 50-200-300 km sec$^{-1}$ and N$_{\mathrm {HI}} \sim10^{21}$ cm$^{-2}$, the Ly$\alpha$ peak shift is even larger than 1000 km sec$^{-1}$. Variations of 500 km sec$^{-1}$ in the shift of the Ly$\alpha$ peak could be related to a $\Delta$log(N$_{\mathrm {HI}}$) = 2. At VUDS resolution, the strongest variations in the location of the Ly$\alpha$ peak shift are therefore probably due to variations in N$_{\mathrm {HI}}$. 
The profiles in Fig. \ref{fig:AnnevsVNHI} are shown for stellar reddening, E(B-V)=0.0. The inclusion of dust in the simulation would reduce the entire Ly$\alpha$ flux, preserving the way the Ly$\alpha$ profile changes when N$_{\mathrm {HI}}$ and V$_{\mathrm {exp}}$ vary. 

Our data show that for the subsamples with $\Delta$v smaller than 100 km sec$^{-1}$ and Ly$\alpha$ peak shifts larger than 300 km sec$^{-1}$, Ext(Ly$\alpha$-C) is larger than 5 kpc.
%
These values can be explained with N$_{\mathrm {HI}}$ = 10$^{20}$-10$^{21}$ cm$^{-2}$ in the framework of the shell model. The scattering of the gas with this N$_{\mathrm {HI}}$ would favour low EW(Ly$\alpha$). 
%
On the other hand, for the subsample with $\Delta$v larger than 300 km sec$^{-1}$ and Ly$\alpha$ peak shifts smaller than 300 km sec$^{-1}$, Ext(Ly$\alpha$-C) is less than 5 kpc. This is consistent with a N$_{\mathrm {HI}} <$ 10$^{20}$ cm$^{-2}$ gas.
%
%
In this case we could expect a large EW(Ly$\alpha$).

If kinematics alone was the main cause of the three Ly$\alpha$ properties, we would note a large Ly$\alpha$ peak shift and Ly$\alpha$ spatial extension in the cases of large outflow velocities. This is not observed in our data. Therefore, the HI column density has a key role in shaping the Ly$\alpha$ emission.

In Fig. \ref{fig:EWExtLyapeak} and \ref{fig:parprop}, we show that extended Ly$\alpha$ spatial profiles are most likely to be produced by scattered light. They could also be due to the ionization of the external shell of the circum-galactic medium by the UV background, which recombines and emits Ly$\alpha$ in situ as fluorescence \citep[e.g.][]{Kollmeier2010, Cantalupo2012}. Another scenario is that of the gravitational cooling of the gas falling onto galaxies \citep[e.g.][]{Rosdahl2012}, which may also lead to some level of scattering of the Ly$\alpha$ photons on their way out. This happens since the Ly$\alpha$ emissivity is proportional to the density, i.e. higher at the center. The scattering process naturally links the spatial extent with the shift of the Ly$\alpha$ peak and the small EW(Ly$\alpha$). The fluorescence and gravitational cooling scenarios can explain any Ly$\alpha$ spatial extension, but does not explain the correlation with the shift of the peak 
nor with the equivalent width. 

Here, we considered the interpretation of the data from homogeneous shell models, but it remains to be studied if other geometries would also show the same behaviour. According to \citet{Verhamme2015}, a scattering medium with holes would produce the direct escape of Ly$\alpha$ photons and hence lines at systemic redshift. 
Our data do not seem to favour the presence of holes. According to \citet{Gronke2016}, clumpy outflows decrease the effective optical depth seen by the Ly$\alpha$ photons compared to the ``real'' column density seen by non-resonant photons along the line of sight, and lead to a small peak shift and boosted escape fraction in comparison to a homogeneous medium of the same outflow velocity and same column density. It remains to be studied if a correlation with the spatial extension, as observed in Fig. \ref{fig:EWExtLyapeak}, would be conserved by clumpy geometries.

%
The galaxies in our sample characterized by a large N$_{\mathrm {HI}}$ (N$_{\mathrm {HI}} \simeq$ 10$^{20}$ cm$^{-2}$) would be those with $\Delta$v that is consistent with a static medium, which is bright in UV,  faint in Ly$\alpha$ with C$_{UV}<2.4$, localized in overdense regions, and 
that has most extended Ly$\alpha$ spatial profiles. 
This means that Ly$\alpha$ properties are clearly related to some galactic properties and so that Ly$\alpha$ photons are mainly produced inside the ISM. 
We find that concentration has the strongest impact on kinematics and that there is a clear (anti-)correlation between Ly$\alpha$ and rest-frame UV concentration. 

Results in the literature \citep[e.g.][]{V:2008,Hashimoto2015} have shown that the star-forming galaxies that are bright in UV, faint in Ly$\alpha$, and more extended in the rest-frame UV are also more massive. It is not surprising that they could also be characterized by large HI amount and possibly N$_{\mathrm {HI}}$. In addition to this, we find that Ly$\alpha$ emitters are characterized by outflow velocities of 200-500 km sec$^{-2}$ and Ly$\alpha$ peak shift of 100-300 km sec$^{-2}$. This is in agreement with the findings in \citet{Verhamme2015}, in which they explained that the shift of the Ly$\alpha$ red peak can be related but is not directly proportional to the velocity of the outflowing shell. 
%
%
Rest-frame UV compact sources also tend to be characterized (on average) by outflows of large velocities, which could be favoured by the small size under the same star formation rate activity. 

%
%

An important consequence of the interpretation of our results is that it is possible to use Ly$\alpha$ emission to trace the HI gas and hence infer about the processes that can regulate the distribution and kinematics of the gas into and out of galaxies during their evolution. If Ly$\alpha$ emitters are characterized by strong outflows and low HI column densities, it is possible that they are seen in moments in which they are experiencing very active phases of star formation that consume the gas very fast (see also A17). 

It is, also, interesting and could have implication for the epoch of the re-ionization \citep{Stark2015}, that galaxies with larger EW(Ly$\alpha$) and EW(CIII]1908) are also those with smaller Ext(Ly$\alpha$-C) and smaller Ly$\alpha$ peak shifts on average. 
In fact, at $z\simeq7$ Ly$\alpha$ peak shifts of the order of 200 km sec$^{-1}$ 
could be interpreted as the fact that the intergalactic medium is still ionized \citep[see also][]{Pentericci2016}.

\subsection{Comparison with previous studies}
\label{sec:literature}

The role of kinematics and HI column density in conditioning the escape of Ly$\alpha$ photons and in shaping the Ly$\alpha$ emission line was studied in detail in high-resolution spectra. \citet{Hashimoto2015} found that the $z\simeq2.2$ Ly$\alpha$ emitters 
of their sample with Ly$\alpha$ peaks close to the systemic redshift had properties consistent with an N$_{\mathrm {HI}} <$ 10$^{19}$ cm$^{-2}$. 
Moreover, they observed an anti-correlation between N$_{\mathrm {HI}}$ and Ly$\alpha$ peak shift. 
\citet{Erb2014,Trainor2015} proposed that the shape of the Ly$\alpha$ line at $z\sim2-3$ is mainly due to the gas column density and covering fraction, while the kinematics would mostly affect the line wings \citep[see also][]{Steidel2010}. Also, they found no evidence that the shift of the Ly$\alpha$ peak is directly proportional to the outflow velocity  
and that the Ly$\alpha$ peak shift anti-correlates with EW(Ly$\alpha$) \citep[see also][]{Shibuya2013}. These observations are consistent with our results.

At low redshift, the escape of Ly$\alpha$ photons was extensively studied using the Cosmic Origins Spectrograph (COS) on board the Hubble Space Telescope (HST). 
Observations of local UV-selected galaxies from the Ly$\alpha$ Reference Sample \citep[LARS;][]{GO2014} and Green Pea galaxies \citep{Cardamone2009}, a sample of compact, metal-poor starbursts at $z\simeq0.1-0.3$ that are good analogues of high-redshift LAEs \citep{Amorin2010, Amorin2012, Jaskot2013}, have shown that outflows are a necessary but not sufficient condition for the Ly$\alpha$ photons to escape star-forming regions \citep{Rivera-Thorsen2015,Henry2015}. \citet{YangMalhotra2016} discovered a significant anti-correlation between Ly$\alpha$ escape fraction and N$_{\mathrm {HI}}$ by means of fitting the Ly$\alpha$ emission line profile of Green Pea spectra with shell models. 
\citet{YangMalhotra2017} extracted the Ly$\alpha$ spatial extension from the 2D COS spectra and showed a trend for which large escape fraction values correspond to compact Ly$\alpha$ morphology. Our results at high redshift are in agreement with these trends.

The measurements for the subsamples based on density can tell us that large equivalent width Ly$\alpha$ emitting galaxies tend to prefer underdense regions, where they also seem to support larger outflow velocities. Galaxies with lower EW(Ly$\alpha$) would be mainly located in overdense regions and would be characterized by larger scale Ly$\alpha$ emissions. We speculate that galaxy-galaxy interactions in overdense regions could increase the column density of the HI gas in the close circum-galactic medium. That gas would be able to scatter Ly$\alpha$ photons and reproduce our results. \citet{Matsuda2012,Momose2016} showed that the size of the Ly$\alpha$ spatial extension is indeed proportional to the density. \citet{Gronke2016} explored the effect of cold HI and hot HII gas on the Ly$\alpha$ emission. According to their model, the suppression of the flux in the line center and the extended Ly$\alpha$ spatial profile could also be produced by the properties of the gas in the circum-galactic medium. \citet{Cooke2013} identified Ly$\alpha$ as an environment diagnostic. They found that the star-forming galaxies tend to show Ly$\alpha$ in absorption in massive, group-like halos, while Ly$\alpha$ in emission tends to be observed for galaxies in the outskirts of massive halos, maybe on top of filamentary structures. Our results are also in agreement with these observations.

To summarize, our results are consistent with a scenario in which the ISM kinematics helps the escape of Ly$\alpha$ photons in the sense that the EW(Ly$\alpha$) could be larger than 40 {\AA}, the Ly$\alpha$ peak is red-shifted with respect to the systemic redshift, and  the Ly$\alpha$ spatial extension is larger than the continuum. However, an N$_{\mathrm {HI}}>$ 10$^{20}$ cm$^{-2}$ can produce Ly$\alpha$ peak shifts larger than 300 km sec$^{-1}$ and could allow Ext(Ly$\alpha$-C) to be up to 7 kpc, 
especially in a static ISM. There is evidence that kinematics and N$_{\mathrm {HI}}$ both contribute to explain the trends we see. However, the outflow velocity seems to mainly contribute to EW(Ly$\alpha$) and the outflow velocity, together with N$_{\mathrm {HI}}$, also explain the Ly$\alpha$ line shape and large-scale spatial extension. Combined with recent results in the literature, the low NHI together with low metallicity and high-ionization capability would make a galaxy a strong Ly$\alpha$ emitter \citep{Trainor2016}. 
\section{Conclusions}
\label{sec:Conclusions}

In this work, we have considered a sample of Ly$\alpha$ emitting galaxies from VUDS. We chose the range of redshift of $2<z<4$ to be able to study the Ly$\alpha$ emission line and the systemic-redshift 
sensitive features within one single rest-frame UV spectrum (Sec. \ref{sec:Data}). This avoids the wavelength calibration uncertainty that is sometimes related to using optical data to study rest-frame UV lines and NIR spectra to infer $z_{\rm sys}$. We measured the systemic redshift of the individual galaxies from the CIII]1908 emission line and we obtained stacked spectra of subsamples of galaxies based on the spectroscopic and physical properties listed in the first column of Table \ref{tab:sub}. Therefore, the sample of sources we consider is composed of Ly$\alpha+$ CIII] emitters. These sources share the same typical stellar mass and stellar mass density as the typical Ly$\alpha$ emitting galaxies in VUDS and are characterized by larger EW(Ly$\alpha$).

From the stacked spectra, we derived the HI outflow velocities by measuring the velocity offset, $\Delta$v, between the systemic redshift and the mean of the Gaussian best fit of the low-ionization absorption lines (SiII1260 and SiII1526). Also, we studied the spatial extension of the Ly$\alpha$ profile, Ext(Ly$\alpha$-C), from the 2D spectra (Sect. \ref{sec:Method}). 

In Table \ref{tab:sub} and Fig. \ref{fig:Paramsig}, \ref{fig:EWExtLyapeak}, and \ref{fig:parprop}, we summarize the main results of our analysis, which are as follows:

$\bullet$ The typical galaxies in our subsamples have the $\Sigma_{\mathrm {SFR}}$ conditions to support outflows. 

$\bullet$ The subsamples for which we measured large Ly$\alpha$ peak shifts ($> 300$ km sec$^{-1}$), extended Ly$\alpha$ spatial profiles (up to 7 kpc more extended than the UV continuum, based on our definition), and small EW(Ly$\alpha$) are characterized by $\Delta$v(SiII) that is either consistent with a static medium or a medium with an equal contribution of inflows and outflows.

$\bullet$ The subsample of rest-frame UV faint Ly$\alpha$ emitters present less extended Ly$\alpha$ spatial profiles, small Ly$\alpha$ peak shifts, and we measure velocity offsets consistent with outflow velocities of a few hundreds of km sec$^{-1}$. Also they tend to prefer underdense regions. 

$\bullet$ By combining our results with the predictions from the radiative-transfer model we can  observe that, at VUDS resolution, we are able to notice the effect of HI column density in the escape and spatial extension of Ly$\alpha$ photons. 

$\bullet$ From the comparison with the model predictions, we can say that a Ly$\alpha$ peak shift larger than $300$ km sec$^{-1}$ can be observed from an ISM with HI column density more than $10^{20}$ cm$^{-2}$. This value of N$_{\mathrm {HI}}$ would favour an efficient scattering process even in the case of a static medium. With the large N$_{\mathrm {HI}}$ we can also explain the large Ly$\alpha$ extension and the small EW(Ly$\alpha$). 
On the other hand, a large $\Delta$v would imply a Ly$\alpha$ spatial profile that is peaked in the galaxy center (i.e. low values of Ext(Ly$\alpha$-C)), and therefore a large EW(Ly$\alpha$), as in our data.

$\bullet$ 
The sources that are more compact in the rest-frame UV are characterized by larger outflow velocities than the paired sample. The galaxies preferentially located in overdense regions also present larger Ext(Ly$\alpha$-C) values than the paired sample. We speculate that this can be produced by galaxy-galaxy interactions, which could increase the N$_{\mathrm {HI}}$ in the close circum-galactic medium.

$\bullet$ Our results and their interpretation via radiative-transfer models tell us that it is possible to use Ly$\alpha$ to trace the properties of the HI gas. Also, the fact that Ly$\alpha$ emitters are characterized by large $\Delta$v could give hints about their stage of evolution. They could be experiencing short bursts of star formation that push strong outflows. More massive star-forming galaxies could be experiencing more normal phases of star formation, which slowly consume the HI gas, and could be showing extended Ly$\alpha$ spatial profiles favoured by the larger HI column density.

$\bullet$ We have shown in Fig. \ref{fig:EWExtLyapeak} the correlation between Ly$\alpha$ spatial and spectral escape from galaxies. It is expected in case of scattering processes. This implies that the origin of the extended Ly$\alpha$ spatial profiles is the scattered light. 

To confirm the role of kinematics and of HI column density in favouring the escape and distribution of Ly$\alpha$ photons, we will study higher resolution spectra of galaxies from the VANDELS$\footnote{http://vandels.inaf.it/}$ survey. The VANDELS survey aims to deliver very deep VIMOS spectra (at least 20h exposition on source with a maximum of 80h) of bright $2.5 < z < 5.5$ and faint $2.5<z<7$ star-forming galaxies. We will be able to detect systemic-redshift sensitive features and LIS lines with enough signal to noise to study kinematics in individual spectra. The HI amount and distribution can be different in different density environments. Therefore, within VANDELS we will also study Ly$\alpha$ as a function of environment and infer the processes that can regulate the distribution and kinematics of the gas into and out of galaxies during their evolution.



\begin{acknowledgements}
We thank the anonymous referee for giving useful suggestions. We thank Benedetta Vulcani and Bram Venemans for useful discussions. 
This work is supported by funding from the European Research Council Advanced Grant ERC--2010--AdG--268107--EARLY and by INAF Grants PRIN 2010, PRIN 2012, and PICS 2013. This work is based on data products made available at the CESAM data center, Laboratoire d'Astrophysique de Marseille, France. 
R.A. acknowledges support from the ERC Advanced Grant 695671 ''QUENCH'' and I.O. from the Czech Science Foundation grant 17-06217Y.
\end{acknowledgements}

\bibliographystyle{aa}   
\bibliography{biblio}        


\begin{appendix} 
\section{Stacked spectra of the subsamples reported in Table 1.}
The following figures show the stacked spectra of the galaxies of each subsample. Stellar features are indicated with yellow, ISM absorption lines with black dashed lines.
We clearly see the Ly$\alpha$ (blue dashed line) and CIII]1908 (green dashed line) emission lines. We can see the HeII (red dashed line) at the $z_{\rm sys}$ of the stacks. 
We verified that the emission line ratios are consistent with those of SFGs (not AGNs) based on the models in \citet{Feltre2016}.

\begin{figure*}
\includegraphics[width=20cm]{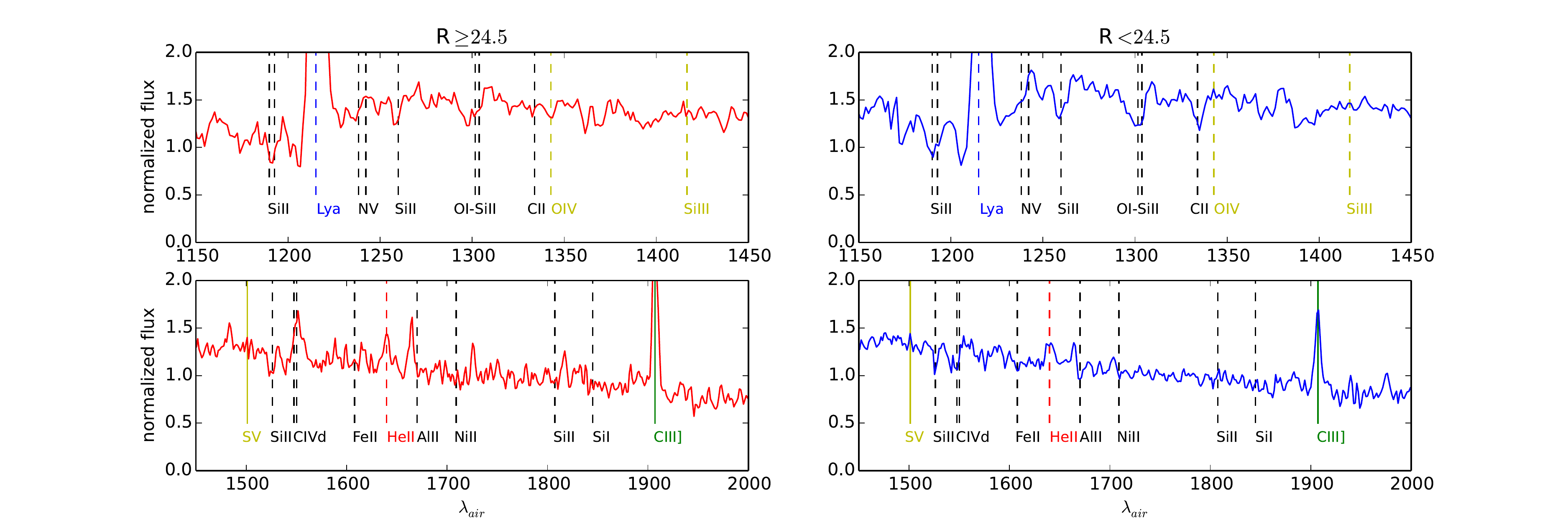}
\includegraphics[width=20cm]{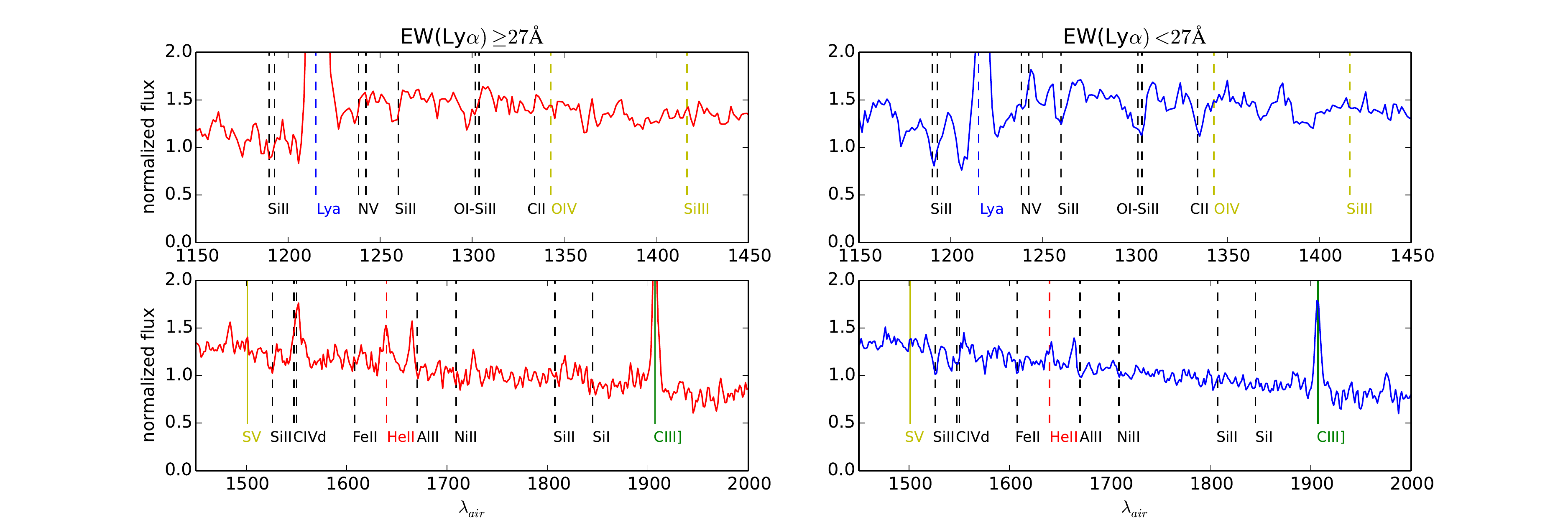}   
\includegraphics[width=20cm]{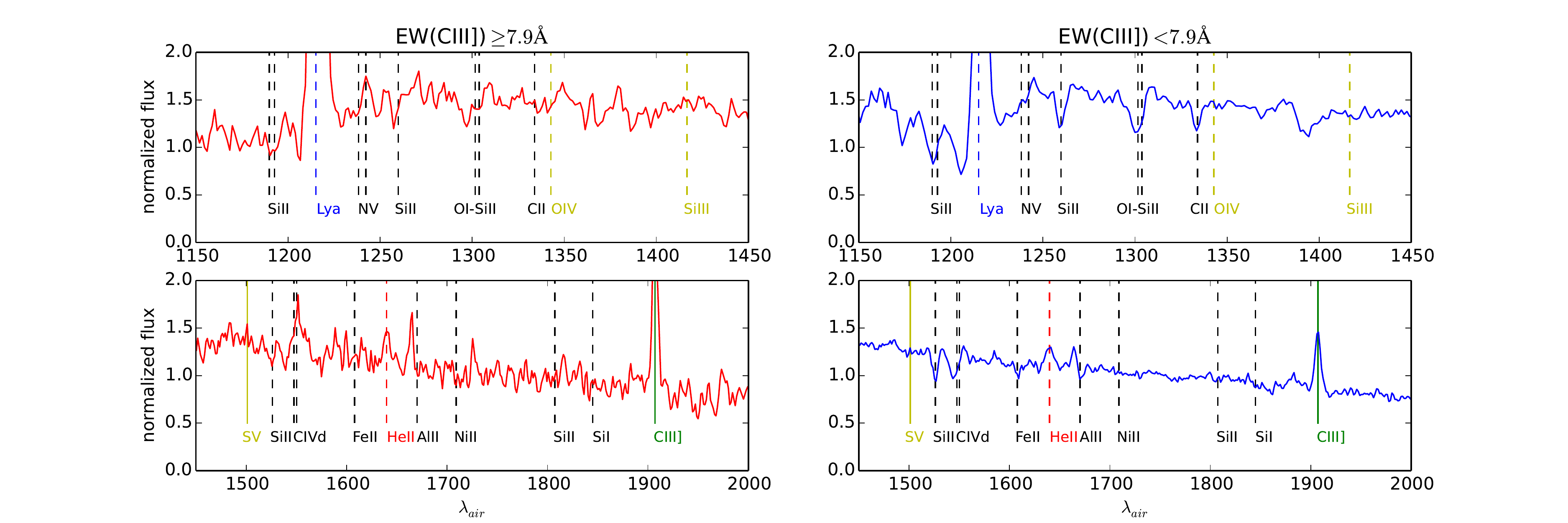}
\caption{Stacked spectra of the subsamples of the galaxies with $R\geq24.5$ and $R<24.5$, EW(Ly$\alpha)\geq27$ {\AA} and EW(Ly$\alpha)<27$ {\AA}, EW(CIII]1908) $\geq7.9$ {\AA} and EW(CIII]19078) $<7.9$ {\AA}. For each subsample we present two panels, one covering the wavelength region with Ly$\alpha$ ($upper$) and the other the wavelength region with CIII]1908 ($lower$).}
\label{stackedspectra1}
\end{figure*}

\begin{figure*}
\includegraphics[width=20cm]{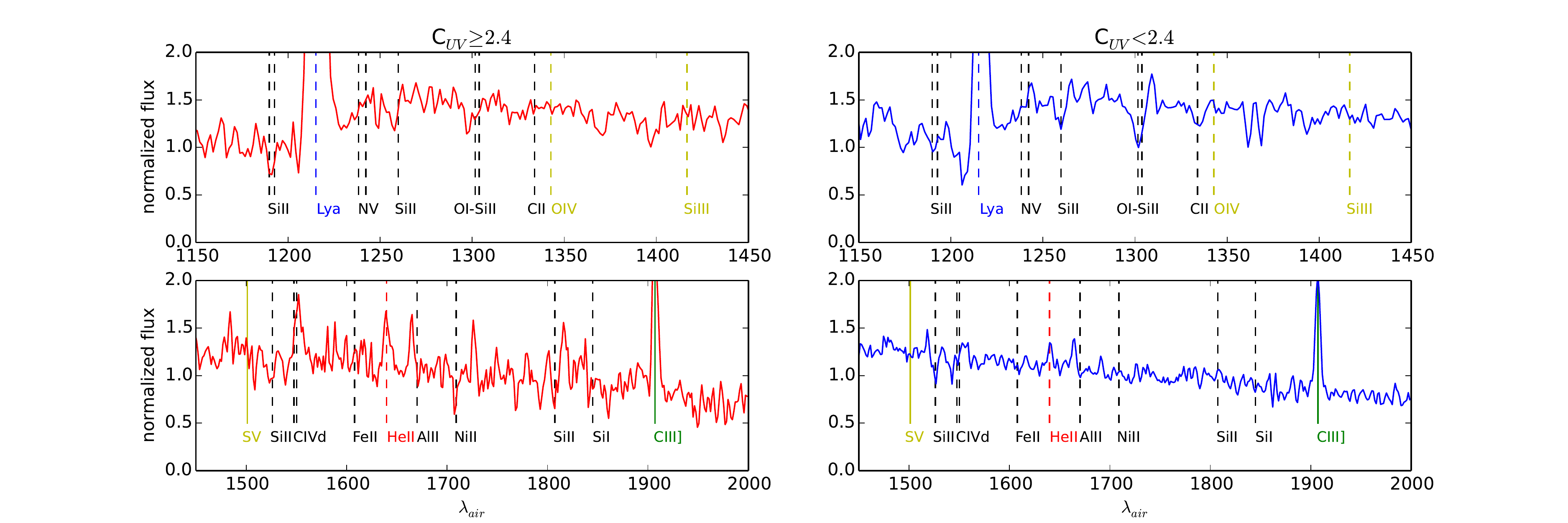} 
\includegraphics[width=20cm]{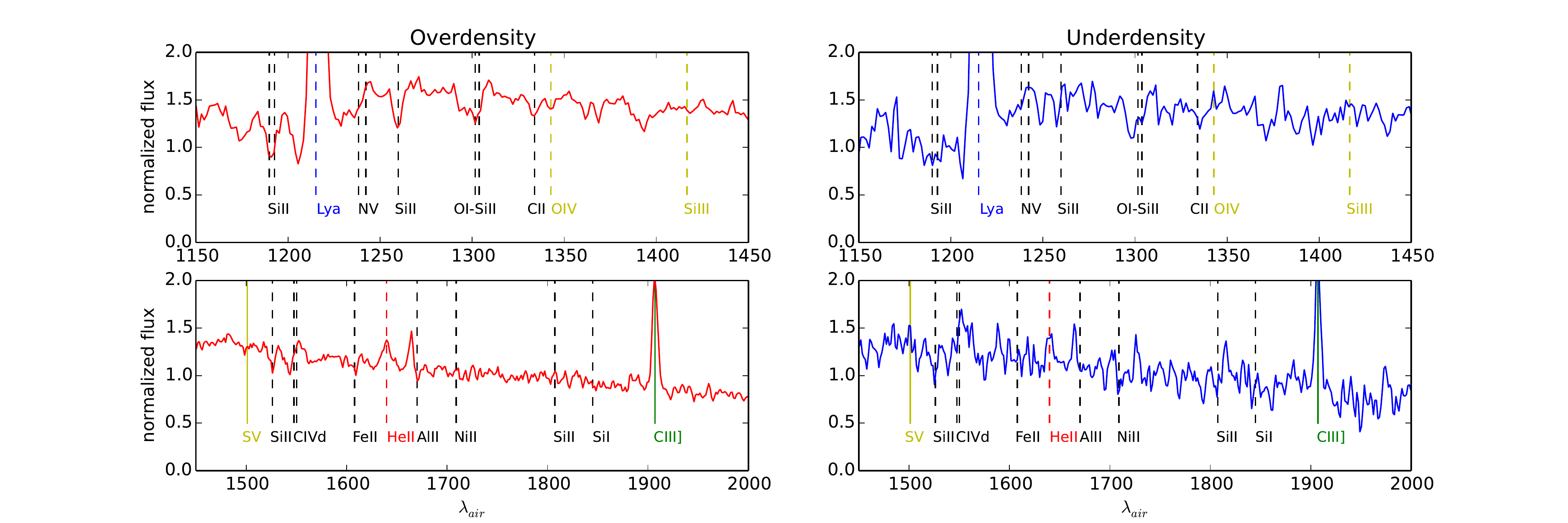}
\includegraphics[width=20cm]{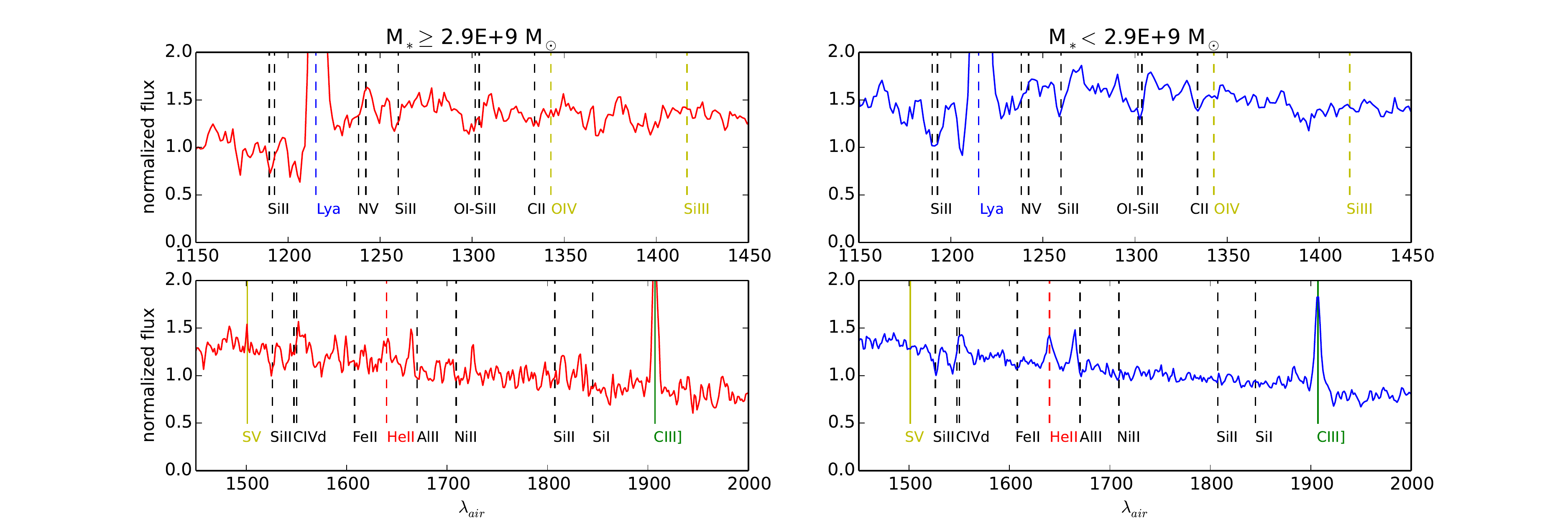}
\caption{Same as Fig. \ref{stackedspectra1} for the subsamples of the galaxies with C$_{UV}\geq 2.4$, C$_{UV}<2.4$, located in over- and underdense regions with M$_{*}\geq2.9$E+9 M$_{\odot}$, and M$_{*}<2.9$E+9 M$_{\odot}$}
\label{stackedspectra2}
\end{figure*}

\begin{figure*}
\includegraphics[width=10cm]{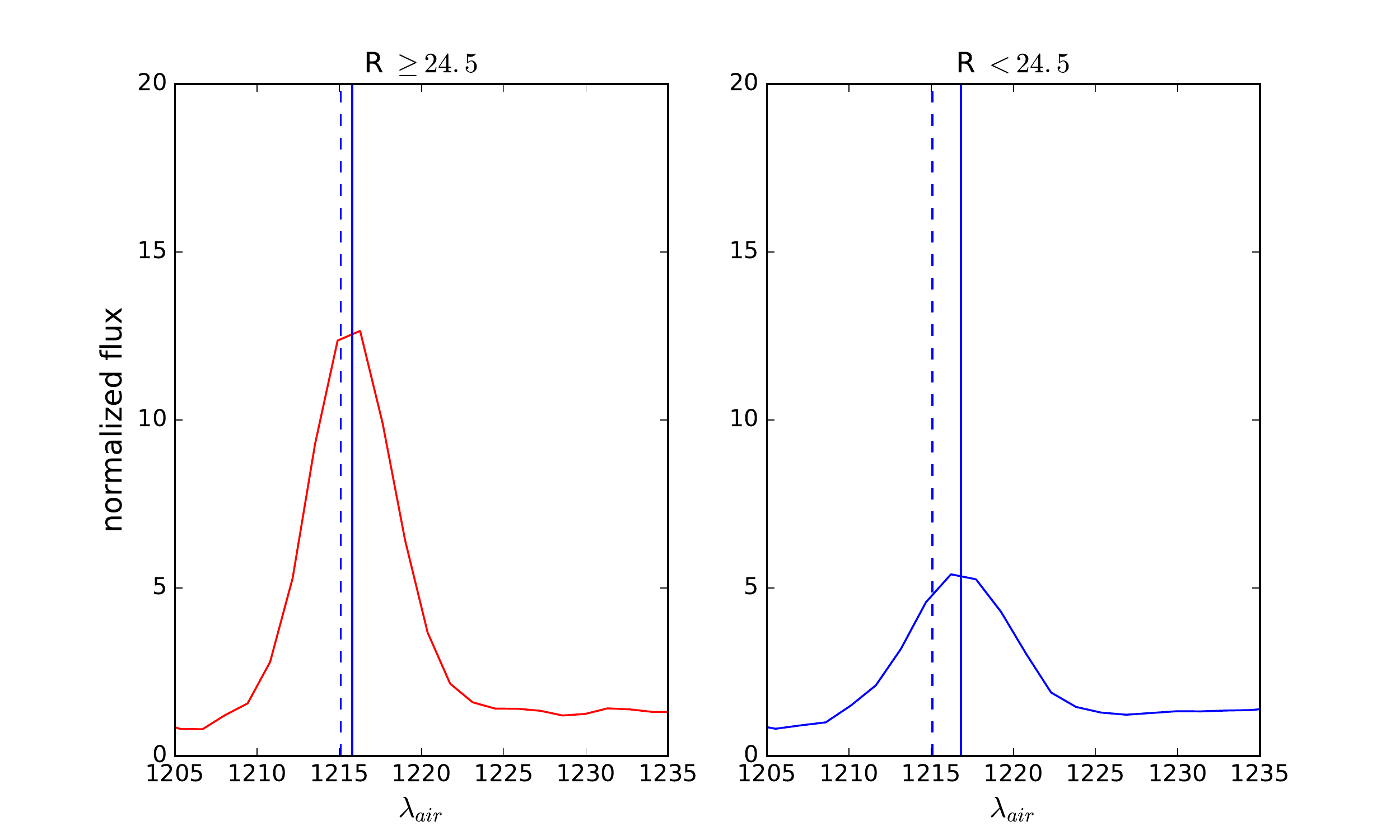}
\includegraphics[width=10cm]{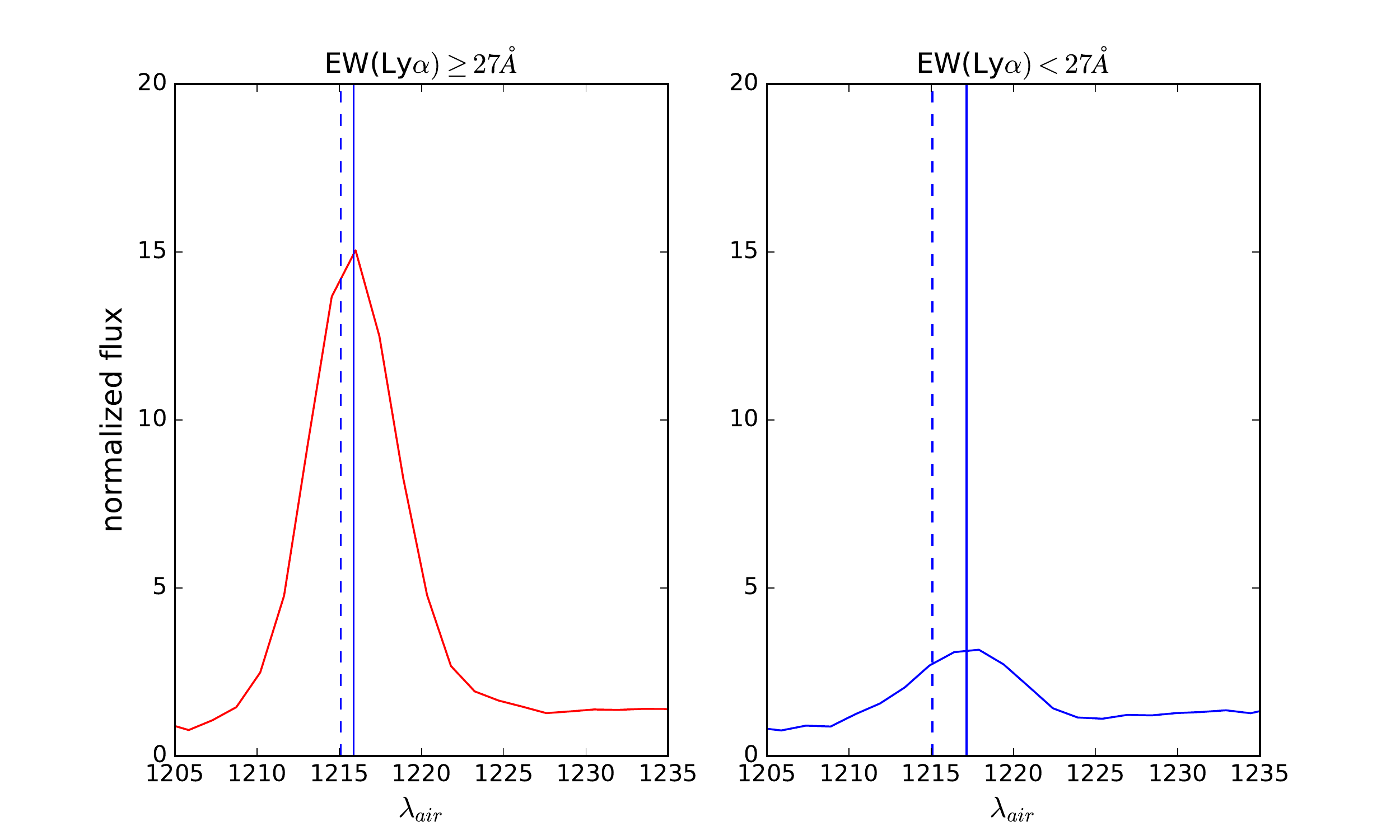}        
\includegraphics[width=10cm]{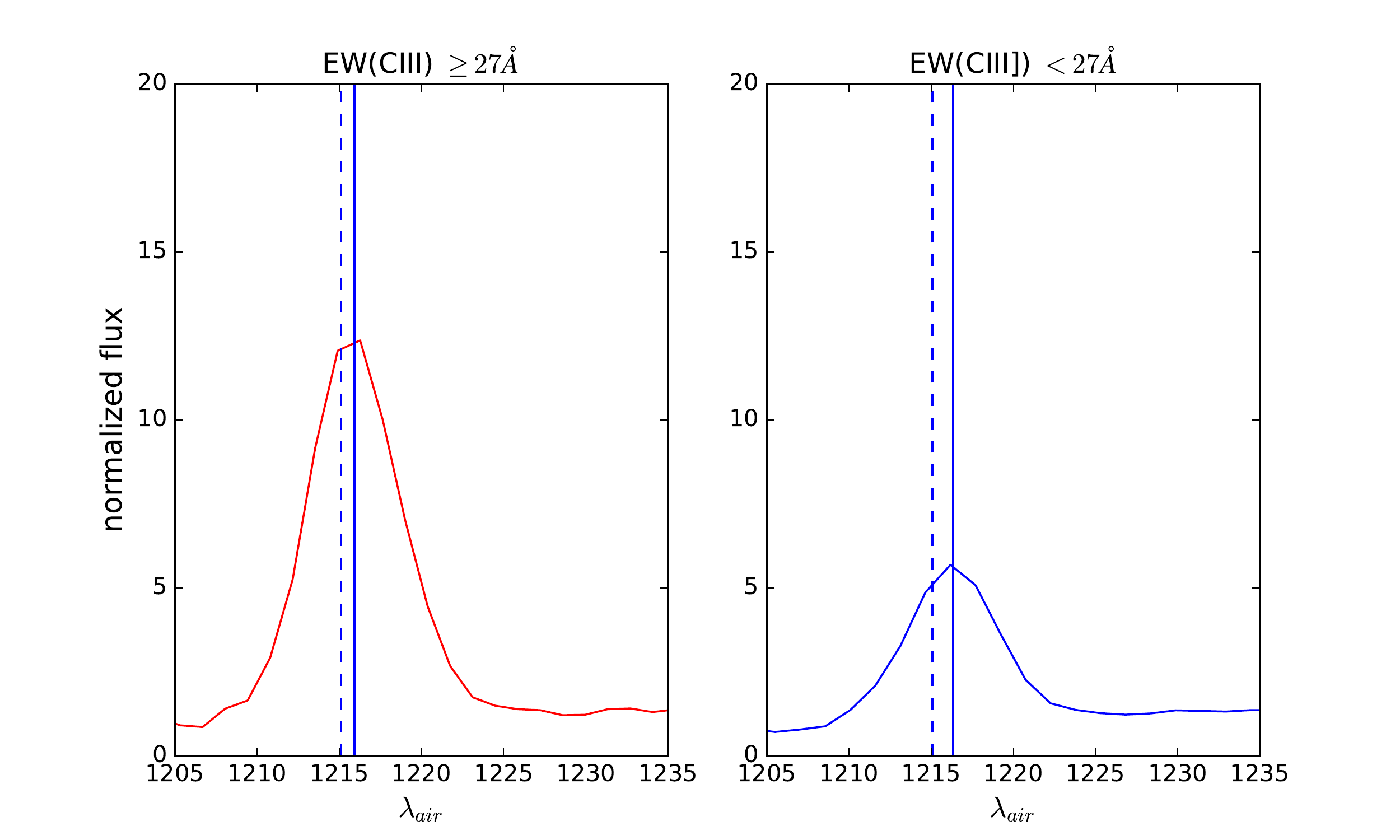}
\includegraphics[width=10cm]{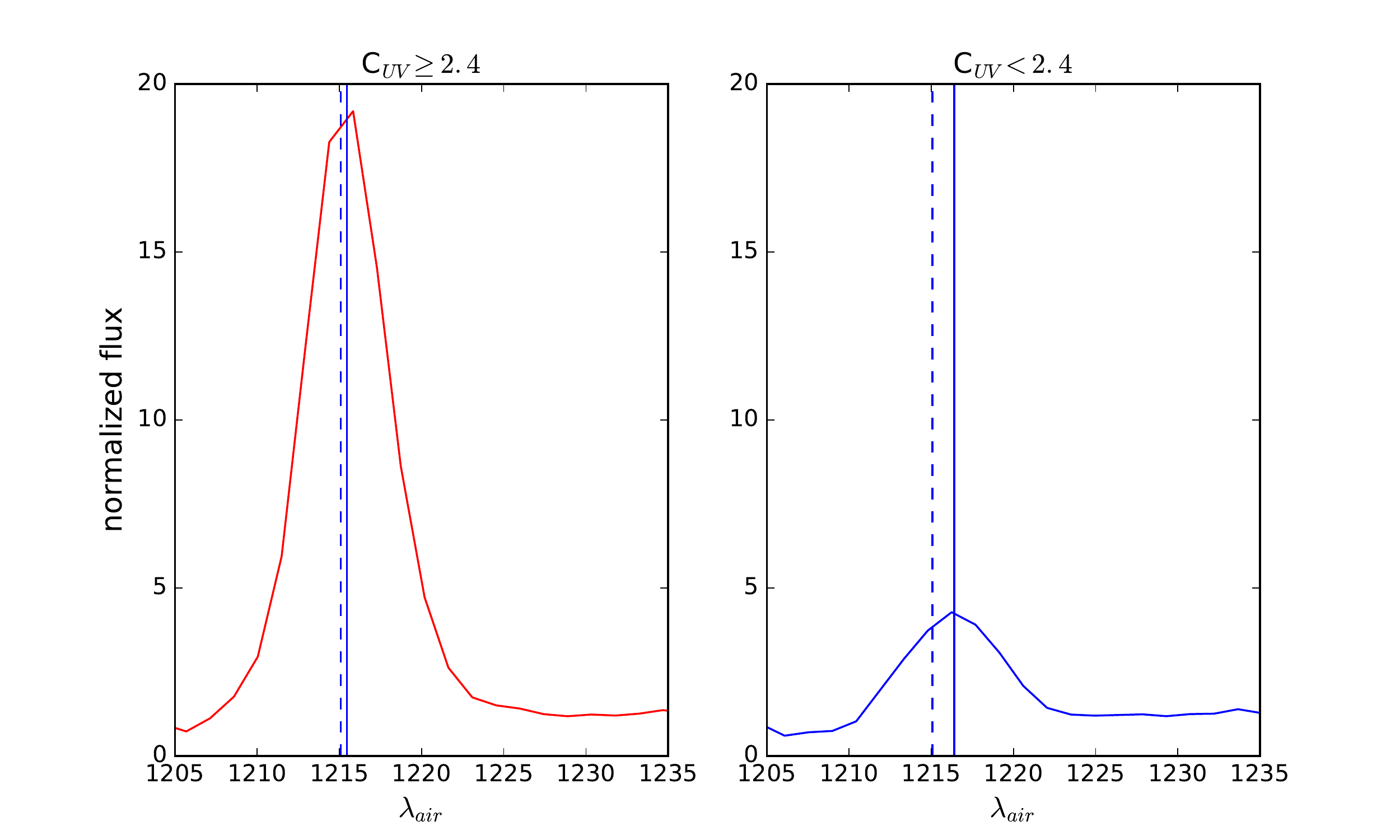}
\includegraphics[width=10cm]{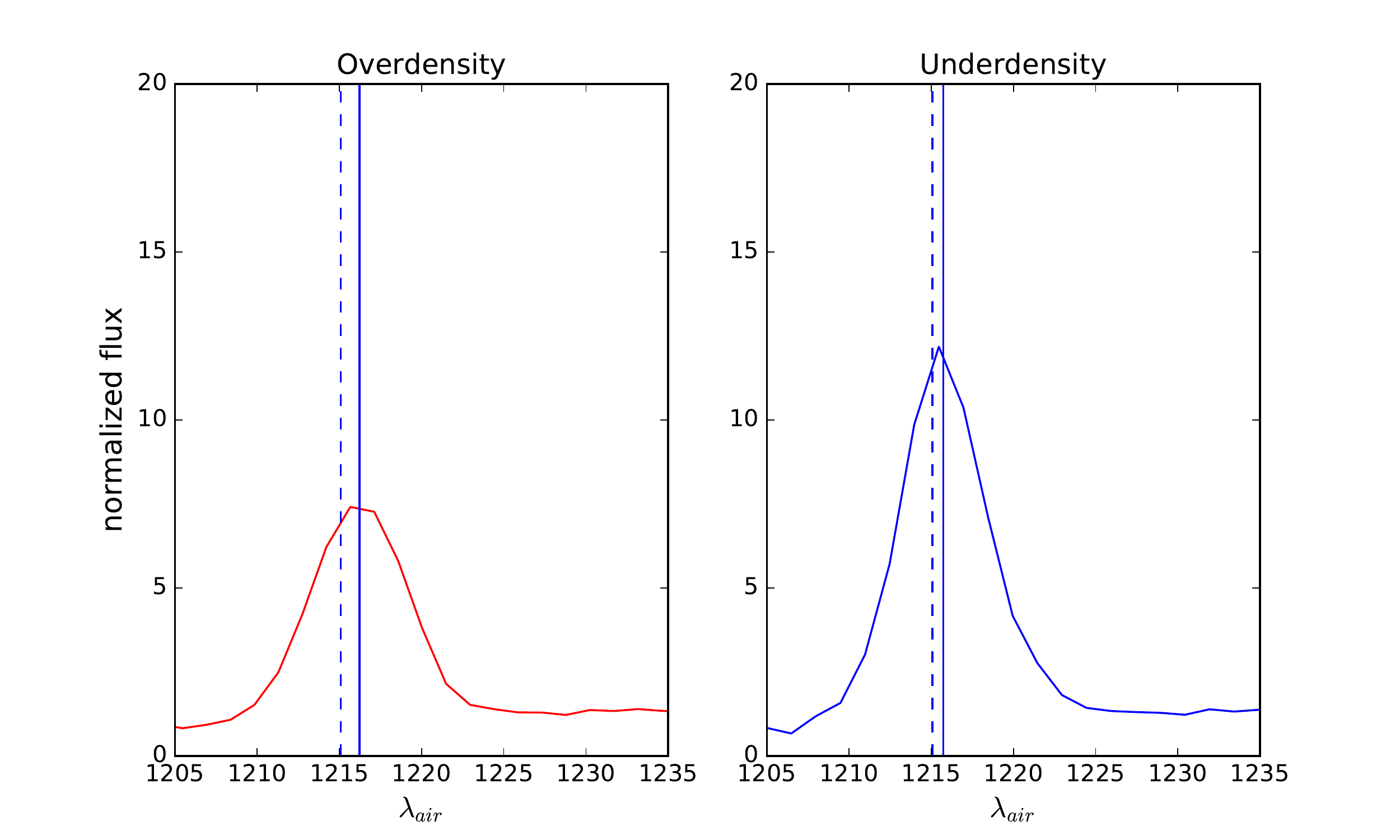}
\includegraphics[width=10cm]{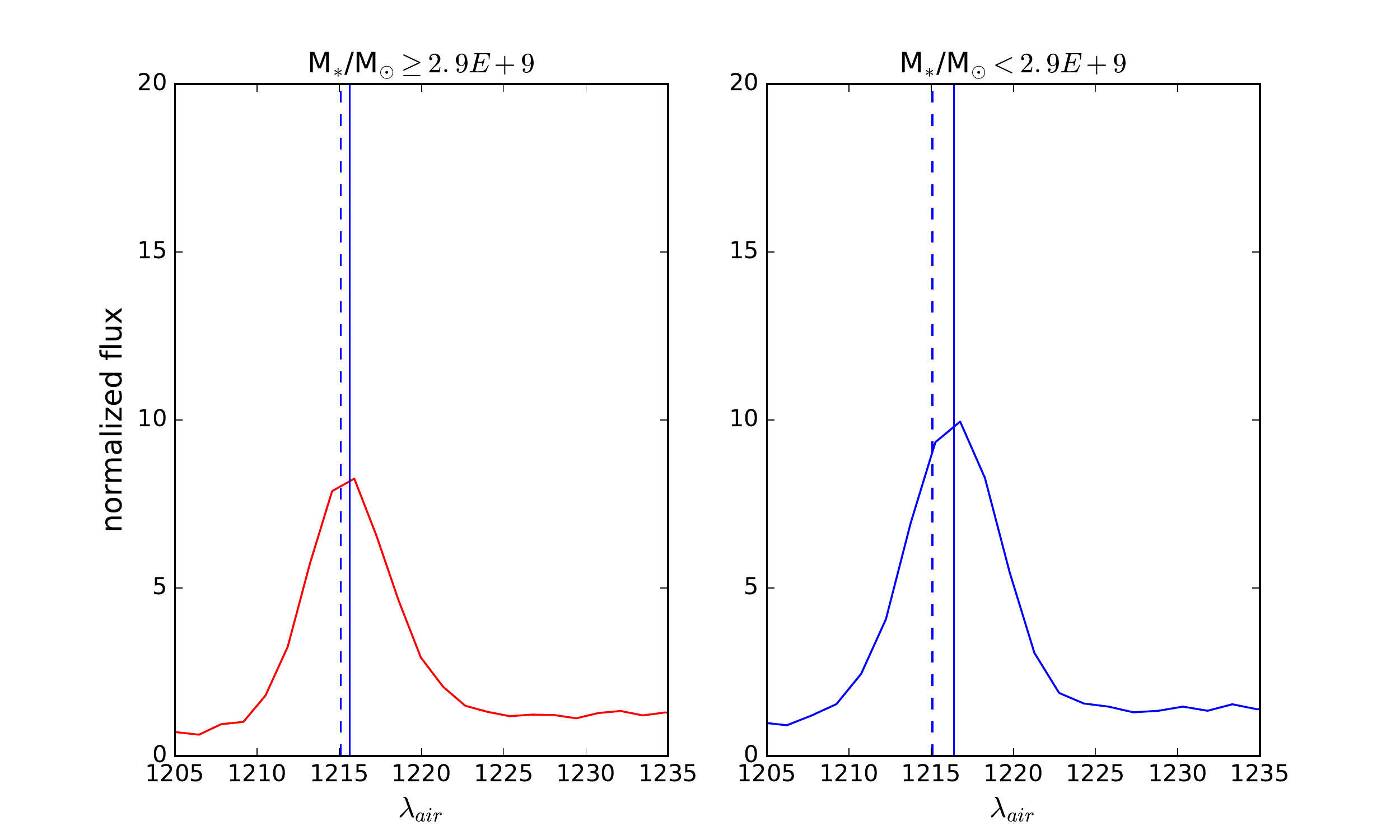}
\caption{Zoom on the Ly$\alpha$ emission line of the stacked spectra in Fig. \ref{stackedspectra1} and \ref{stackedspectra2}.  Vertical dashed lines indicate the rest-frame Ly$\alpha$ wavelength, while the vertical solid lines indicate the wavelength of the red peak.}
\label{stackedonlyLya}
\end{figure*}

Fig. \ref{stackedonlyLya} shows a zoom of the stacked spectra on the Ly$\alpha$ wavelength. The Ly$\alpha$ main peak is redshifted with respect to the systemic redshift for all the subsets studied here.

\end{appendix}
 
\end{document}